\documentclass[preprint,authoryear,11pt]{elsarticle}  
\usepackage{fullpage}
\usepackage{graphicx}
\usepackage{amssymb}
\usepackage{amsmath}
\usepackage{multirow}
\usepackage{lineno}
\usepackage{subfig}

\usepackage{color}
\usepackage{soul}
\usepackage{array}
\usepackage{hyperref}
\usepackage[colorinlistoftodos]{todonotes}
\usepackage[]{algorithm2e}
\usepackage{varwidth}
\graphicspath{{./figures/}}

\journal{}

\begin{document}

\begin{frontmatter}

\title{Investigating the Settling Dynamics of Cohesive Silt Particles With Particle-Resolving Simulations}

\author[1]{Rui Sun} \ead{sunrui@vt.edu}
\author[1]{Heng Xiao} \ead{hengxiao@vt.edu}
\address[1]{Department of Aerospace and Ocean Engineering, Virginia Tech, Blacksburg, Virginia, USA}

\author[2]{Honglei Sun\corref{corhl}} \ead{sunhonglei@zju.edu.cn}
\address[2]{Institute of Disaster Prevention, Zhejiang University, Hangzhou, China}

\cortext[corhl]{Corresponding author. Tel: +86 187 5811 9525}

\begin{abstract}
The settling of cohesive sediment is ubiquitous in aquatic environments, and the study of the
settling process is important for both engineering and environmental reasons. In the settling
process, the silt particles show behaviors that are different from non-cohesive particles due to the
influence of inter-particle cohesive force. For instance, the flocs formed in the settling process
of cohesive silt can loosen the packing, and thus the structural densities of cohesive silt beds are
much smaller than that of non-cohesive sand beds. While it is a consensus that cohesive behaviors
depend on the characteristics of sediment particles (e.g., Bond number, particle size distribution),
little is known about the exact influence of these characteristics on the cohesive behaviors. In
addition, since the cohesive behaviors of the silt are caused by the inter-particle cohesive forces,
the motions of and the contacts among silt particles should be resolved to study these cohesive behaviors
in the settling process.  However, studies of the cohesive behaviors of silt particles in the
settling process based on particle-resolving approach are still lacking. In the present work,
three-dimensional settling process is investigated numerically by using CFD--DEM (Computational
Fluid Dynamics--Discrete Element Method). The inter-particle collision force, the van der Waals
force, and the fluid--particle interaction forces are considered. The numerical model is used to
simulate the hindered settling process of silt based on the experimental setup in the
literature.  The results obtained in the simulations, including the structural densities of the
beds, the characteristic lines, and the particle terminal velocity, are in good agreement with the
experimental observations in the literature. To the authors' knowledge, this is the first time that
the influences of non-dimensional Bond number and particle polydispersity on the structural
densities of silt beds have been investigated separately. The results demonstrate that the cohesive
behavior of silt in the settling process is attributed to both the cohesion among silt particles
themselves and the particle polydispersity. To guide to the macro-scale modeling of cohesive silt
sedimentation, the collision frequency functions obtained in the numerical simulations are also
presented based on the micromechanics of particles. The results obtained by using CFD--DEM indicate
that the binary collision theory over-estimated the particle collision frequency in the flocculation
process at high solid volume fraction.
\end{abstract}


\end{frontmatter}
 

\section{Introduction}

In the natural environment, the settling process occurs ubiquitously, in which the sediment
particles fall continuously towards the seabed, lake floor or river floor to form a loose sediment
layer~\citep{zhao14ig}. The settling of sediment particles can have negative impacts on the
environment. For example, excess sedimentation in waterways can make them too shallow for
navigation; sediment deposition can bury the aquatic habitats and is detrimental to aquatic life. On
the other hand, diminished sedimentation can cause the losses of valuable
wetlands~\citep{mcanally00ca}.  Therefore, the understanding of the settling process is important in
the calculation of sediment budget for engineering, economic and environmental reasons. In
addition, the sedimentation process is found in chemical, mining, pharmaceutical and other
industries due to its importance in the understanding of fluid--solid
separation~\citep{shih87hs,burger01ss,dong09dem,zhao14ig}. 

Since silt is the prevailing sediment fraction in many river systems, the modeling and assessment of
sediment dynamics in these rivers require proper knowledge of the behaviors of
silt~\citep{te15hindered}. The size of silt is larger than clay but smaller than
sand~\citep{krumbein37sbb}. Compared with non-cohesive sand, silt demonstrates some behaviors that
are unique to cohesive particles. Specifically, these behaviors include: (1) the structural
densities (solid volume fractions) of silt beds are smaller than those of sand
beds~\citep{winterwerp04intro,te15hindered}; (2) the trajectories of silt particles are deflected when
the flocs form~\citep{stolzenbach94ef,mazzolani98gi,zhang11lbs}; (3) the critical shear stress to
initiate silt bed motion is larger than that of sand bed~\citep{roberts98eff,lick04init}.  These
behaviors are collectively referred to as ``cohesive behaviors'' hereafter following the
literature~\citep{te15hindered}.  These cohesive behaviors could be caused by the cohesion of silt
particle itself or the inter-mixture of silt particles of different sizes~\citep{te15hindered}. It
is believed that the variation of the characteristics of the sediment particles (e.g., Bond number,
particle size distribution) can significantly influence the cohesive
behaviors~\citep{dong06role,dong09dem}. However, little is known about how exactly the variation in
these characteristics can influence the sedimentation of silt particles.

The modeling of the silt settling process is hindered by the complex dynamics of the particles and
the challenges in predicting the fluid--particle interaction and inter-particle contact. The
traditional empirical approaches based on experimental
measurements~\citep{richardson54sedi,winterwerp04intro,te15hindered} are used to predict the
settling velocity, which is among the most important quantities of interest for the fluid--solid
mixture~\citep{batchelor72hi,hinch77ae,zhao14ig}. At higher particle concentrations, the influence
of return flow and wake formation, the inter-particle collision, and the increased buoyancy effects
are considered in the settling process~\citep{winterwerp02floc}. The influence of flocculation on
cohesive sediment in differential settling process can be also considered by using collision
frequency function to describe inter-particle contact~\citep{krishnappan90ms,winterwerp02floc}. The
traditional approaches are able to predict the settling velocity and the sediment concentration in
the regime where calibration data were obtained, but they heavily rely on empirical
correlations to describe the influence of return flow and inter-particle contact.  Therefore, these
models may lead to large discrepancies in the predictions of the cohesive behavior of silt particles
when outside these regimes because the empirical correlations are not derived directly from first
principles. 

With the growth of available computational resources in the past few decades, the Discrete Element
Method (DEM) has gained
popularity~\citep{jiang93mp,drake01dp,chen11cde,capecelatro13el,schmeeckle2014ns}. The DEM uses
Newton's law of motion to predict individual particle motion, and the detailed packing arrangement
of individual particles in the sediment bed can be captured.  Consequently, the DEM has been used to
predict the distribution of poly-dispersed sediment particle in the aggregates after
sedimentation~\citep{bravo15ns}.  Moreover, the DEM explicitly resolves the interaction among the
sediment particles, and thus the influence of the inter-particle cohesive force on the particle
trajectory can be captured~\citep{dyachenko12ms}.  When the influence of fluid flow (e.g., return
flow, wake formation, pressure) is significant in the sedimentation problems, the fluid flow is
usually resolved using the CFD and the coupled CFD--DEM approach is used. In CFD--DEM approach, the
locally-averaged Navier--Stokes equations are solved for the fluid flow, and the fluid--particle
interaction forces are considered. This coupled approach is not only successful in predicting the
excess pore pressure in the sand bed during the sedimentation
process~\citep{zhao13coupled,zhao14ig}, but can also capture the reduction of the structural
densities in the packed beds of cohesive particles in chemical and mineral engineering
applications~\citep{dong06role, dong09dem}. However, the study of cohesive silt settling and the
cohesive behaviors based on CFD--DEM approach is still lacking.

It has recently been demonstrated that \textit{SediFoam}, a hybrid CFD--DEM solver for
particle-laden flows, is capable of modeling the subaqueous sediment motion in extensive validation
tests. For example, satisfactory predictive performances were observed in the modeling of the
current-induced suspended sediment transport, the generation and migration of sand dune, and the
sediment transport in oscillatory flows~\citep{sun16cfd,sun16sedi,Sun16sm}. In this study, we use
\textit{SediFoam} to study the settling process of cohesive silt particles.  Compared with the
previous study on sand particle settling using CFD--DEM~\citep{zhao14ig}, the size of silt is
smaller and inter-particle cohesion is more significant.  The objectives of the present study are to
(1) demonstrate the capability of CFD--DEM to predict the cohesive behaviors (e.g., decrease of
structural densities) in the settling of cohesive silt, (2) investigate the influence of the
particle characteristics (e.g., Bond number, particle polydispersity) on the structural densities of
the silt beds after sedimentation, and (3) examine the accuracy of the empirical formulas in the
prediction of collision frequency function in the macro-scale modeling of cohesive silt
sedimentation.  While this work focuses on the sedimentation of cohesive silt, the proposed approach
also opens the possibility for first-principle-based simulations of the flocculation and erosion of
cohesive silt.

While earlier works~\citep[e.g.,][]{higashitani01sd,dong06role,dong09dem,zhang11lbs} have used
particle-resolving simulations to study the settling of cohesive particles, the current contribution
is novel in several aspects. First, we investigated the individual effects of Bond number and
particle polydispersity on the bed structural density.  Second, we clearly demonstrated the
capability of CFD-DEM in capturing the effects of flocculation on the settling characteristics and
its temporal evolution for poly-dispersed systems.  Finally, our simulations produced collision
frequency functions, which can provide valuable guidance for future development of such empirical
relations widely used in macro-scale models. 

The rest of the paper is organized as follows. Section 2 introduces the methodology of the present
model, including the mathematical formulation of fluid equations, the particle motion equations, the
fluid--particle interactions, and the modeling of cohesion. The implementation details of the code
and the numerical methods used in the simulations are described in Section 3.  In Section 4, the
results obtained in the numerical simulations are presented. Section 5 discusses the insights gained
from the present results for macro-scale modeling. Finally, Section 6 concludes the paper.

\section{Methodology}
\label{sec:cfddem}

\subsection{Mathematical Model of Particle Motion}
\label{sec:dem}
In CFD--DEM, the modeling of translational and rotational motion of each sediment particle is based
on Newton's second law as the following equations~\citep{cundall79}: 
\begin{subequations}
 \label{eq:newton}
 \begin{align}
  m \frac{d\mathbf{u}}{dt} &
  = \mathbf{f}^{col} + \mathbf{f}^{vdw} + \mathbf{f}^{fp} + m \mathbf{g} \label{eq:newton-v}, \\
  I \frac{d\boldsymbol{\Psi}}{dt} &
  = \mathbf{T}^{col} + \mathbf{T}^{fp} \label{eq:newton-w},
 \end{align}
\end{subequations}
where $m$ is the mass of particle; \( \mathbf{u} \) is particle velocity; $t$ is time;
\(\mathbf{f}^{col} \), \(\mathbf{f}^{vdw}\), \(\mathbf{f}^{fp}\) are inter-particle collision force,
van der Waals force, fluid--particle interaction forces, respectively; \(\mathbf{g}\) is the
gravitational acceleration. Similarly, \(I\) and \(\boldsymbol{\Psi}\) are angular moment of inertia
and angular velocity of the particle; \(\mathbf{T}^{col}\) and \(\mathbf{T}^{fp}\) are the torques
due to inter-particle collision and fluid--particle interactions, respectively. To compute the
collision forces and torques, the particles are modeled as soft spheres with inter-particle contact
represented by an elastic spring and a viscous dashpot. The detailed description of the collision
model in \textit{SediFoam} and the extensive validations tests with respective to the experimental
data are available in~\cite{gupta15vv}.

\subsection{Cohesion Model}
\label{sec:cohesion}
In the present simulations, the van der Waals force is implemented and used to model the cohesion
among sediment particles~\citep{israelachvili11is}:
\begin{equation}
  \mathbf{f}^{vdw}_{ij} =  -\frac{H_a}{6}
  \frac{64r_i^3 r_j^3\left(h + r_i + r_j\right)}
  {\left( h^2 + 2r_ih + 2r_jh \right)^2\left( h^2 + 2r_ih + 2r_jh + 4r_ir_j \right)^2}
  \mathbf{n}_{ij},
  \label{eq:vdw}
\end{equation}
where $\mathbf{f}^{vdw}_{ij}$ is the van der Waals force between particle $i$ and $j$; $H_a$ is the
Hamaker coefficient; $r_i$ and $r_j$ are the radii of particle $i$ and $j$, respectively; $h$ is the
separation distance between the two particles $i$ and $j$; $\mathbf{n}_{ij}$ is the unit vector from
particle $j$ to particle $i$. A minimum separation distance $h_{min}$ is applied to avoid
singularity when $h$ equals to zero~\citep{yang00cs,dong09dem}. Since the minimum separation
distance is much smaller than particle size ($h \ll r$), Eq.~(\ref{eq:vdw}) reduces to:
\begin{equation}
  \mathbf{f}^{vdw}_{max,ij} =  -\frac{H_a r}{12h_{min}^2}\mathbf{n}_{ij}.
  \label{eq:vdwMax}
\end{equation}
The influence of cohesive force to the silt particle is represented by using the dimensionless Bond
number $Bo$, which is defined as the ratio of the maximum cohesive force $\mathbf{f}^{vdw}_{max}$
and particle weight~\citep{clift78bdp}:
\begin{equation}
  Bo = \left|\frac{\mathbf{f}^{vdw}_{max}}{m\mathbf{g}}\right|
  = \frac{H_a r}{12 h_{min}^2 mg},
  \label{eq:bond}
\end{equation}
where $g$ is the magnitude of gravitational acceleration. It should be noted that both van der Waals
force and electrostatic repulsive force act among particles in DLVO theory for particle
cohesion~\citep{verwey99theory,israelachvili11is}. However, the electrostatic force does not
significantly influence particle cohesion in the settling process and thus is not considered in the
present simulation for the sake of simplicity. This simplification is consistent with these made in
the literature~\citep{higashitani01sd,zhang11lbs}.

\subsection{Locally-Averaged Navier--Stokes Equations for Fluids}
\label{sec:lans}
Locally-averaged incompressible Navier--Stokes equations are used to describe the fluid flow.
Assuming constant fluid density \(\rho_f\), the governing equations for the fluid
are~\citep{anderson67,kafui02}:
\begin{subequations}
 \label{eq:NS}
 \begin{align}
  \nabla \cdot \left(\varepsilon_s \mathbf{U}_s + {\varepsilon_f \mathbf{U}_f}\right) &
  = 0 , \label{eq:NS-cont} \\
  \frac{\partial \left(\varepsilon_f \mathbf{U}_f \right)}{\partial t} + \nabla \cdot \left(\varepsilon_f \mathbf{U}_f \mathbf{U}_f\right) &
  = \frac{1}{\rho_f} \left( - \nabla p + \varepsilon_f \nabla \cdot \boldsymbol{\mathcal{R}} + \varepsilon_f \rho_f \mathbf{g} + \mathbf{F}^{fp}\right), \label{eq:NS-mom}
 \end{align}
\end{subequations}
where \(\varepsilon_s\) is the solid volume fraction; \( \varepsilon_f = 1 - \varepsilon_s \) is the
fluid volume fraction; \( \mathbf{U}_f \) is the fluid velocity. The terms on the right hand side of
the momentum equation are: pressure gradient \(\nabla p\), divergence of the stress tensor \(
\boldsymbol{\mathcal{R}} \), gravity, and fluid--particle interaction forces, respectively. In the
present study, only the viscous stress is considered because the fluid Reynolds number is small. The
Eulerian fields $\varepsilon_s$, $\mathbf{U}_s$, and $\mathbf{F}^{fp}$ in Eq.~(\ref{eq:NS}) are
obtained by averaging the Lagrangian information of particles. 

\subsection{Fluid--Particle Interactions}
\label{sec:fpi}
The fluid--particle interaction forces \(\mathbf{f}^{fp}\) in Eq.~(\ref{eq:newton-v}) consist of
buoyancy \( \mathbf{f}^{buoy} \), drag \( \mathbf{f}^{drag} \), lift force \(\mathbf{f}^{lift}\),
and added mass force \(\mathbf{f}^{am}\).  Although the lift force and the added mass force are
not considered in the validation tests of fluidized bed simulation~\citep{gupta15vv}, they are
important in the simulation of sediment transport.

The drag force model proposed by~\cite{mfix93} is applied in the present simulations. The drag on
an individual component sphere $i$ is formulated as:
\begin{equation}
  \mathbf{f}^{drag}_i = \frac{V_{p,i}}{\varepsilon_{f, i} \varepsilon_{s, i}} \beta_i \left(
  \mathbf{U}_{f, i}  - \mathbf{u}_{p,i} \right),
  \label{eq:particleDrag}
\end{equation}
where \( V_{p, i} \) and \( \mathbf{u}_{p, i} \) are the volume and the velocity of particle $i$,
respectively; \( \mathbf{U}_{f, i} \), \( \varepsilon_{f, i} \), and \( \varepsilon_{f, i} \) are the
fluid velocity, solid volume fraction, and fluid volume fraction interpolated to the center of
particle $i$, respectively; \( \beta_{i} \) is the drag correlation coefficient which accounts for
the presence of other particles. The $\beta_i$ value in the present study is based on~\cite{mfix93}:
\begin{equation}
  \beta_i = \frac{3}{4}\frac{C_{d,i}}{V^2_{r,i}} \frac{\rho_f |\mathbf{U}_{f,i} -
  \mathbf{u}_{p,i}|}{d_{p,i}}\varepsilon_{f, i} \varepsilon_{s, i} \mathrm{, \quad with \quad}
  C_{d,i} = \left( 0.63+0.48\sqrt{\Gamma_{r,i}/\mathrm{Re_{p,i}}} \right),
  \label{eq:beta-i}
\end{equation}
where $C_{d,i}$ is the drag coefficient of particle $i$; $d_{p,i}$ is the diameter of particle $i$;
the particle velocity Reynolds number $\mathrm{Re_{p,i}}$ is defined as:
\begin{equation}
  \mathrm{Re_{p,i}} = \rho_s d_{p,i} |\mathbf{U}_{f,i} - \mathbf{u}_{p,i}|/\mu,
  \label{eq:p-re}
\end{equation}
where $\rho_s$ is the density of solid particle, $d_{p,i}$ is the diameter of particle $i$, $\mu$ is
the dynamic viscosity of fluid flow; the $\Gamma_{r,i}$ is the correlation term for the $i$-th
particle:
\begin{equation}
  \Gamma_{r,i} = 0.5\left( A_{1,i} - 0.06\mathrm{Re_{p,i}}+\sqrt{(0.06\mathrm{Re_{p,i}})^2 +
  0.12\mathrm{Re_{p,i}}(2A_{2,i} - A_{1,i})+A_{1,i}^2} \right),
  \label{eq:drag-vr}
\end{equation}
with
\begin{equation}
  A_{1,i} = \varepsilon_{f,i}^{4.14}, \quad
  A_{2,i} =
  \begin{cases}
  0.8\varepsilon_{f,i}^{1.28} & \quad \text{if } \varepsilon_{f,i} \le 0.85, \\
  \varepsilon_{f,i}^{2.65}    & \quad \text{if } \varepsilon_{f,i} > 0.85.\\
  \end{cases}
  \label{eq:drag-A}
\end{equation}
In addition to drag, the lift force on a spherical component particle is modeled
as~\citep{saffman65th,rijn84se1,zhu07dps}:
\begin{equation}
  \mathbf{f}_{i}^{lift} = C_{l} (\rho_f \mu)^{0.5} d_{p,i}^{2} \left( \mathbf{U}_{f,i} -
  \mathbf{u}_{p,i} \right) \mathbf{\times}
  \frac{\boldsymbol{\omega}_i}{|\boldsymbol{\omega}_i|^{0.5}},
  \label{eq:particleLift}
  \end{equation}
where $C_{l} = 1.6$ is the lift coefficient, $\boldsymbol{\omega}_i = \nabla \times
\mathbf{U}_{f,i}$ is the curl of flow velocity interpolated to the center of particle $i$, and
$\mathbf{\times}$ indicates the cross product. The added mass force is considered important because
the densities of the carrier and disperse phases are comparable in sediment transport applications.
This is modeled as:
\begin{equation}
  \mathbf{f}_{i}^{am} = C_{am} \rho_f V_{p,i} \left( \frac{\mathrm{D}\mathbf{U}_{f,i}}{\mathrm{D}t}
  - \frac{\mathrm{d}\mathbf{u}_{p,i}}{\mathrm{d}t} \right),
  \label{eq:particleAddedMass}
  \end{equation}
where $C_{am} = 0.5$ is the coefficient of added mass;  $\frac{\mathrm{D}
\mathbf{U}_{f,i}}{\mathrm{D}t}$ represents the material derivative of fluid velocity interpolated to
the center of particle $i$. 

\section{Implementations and Numerical Methods}
\label{sec:num-method}
The hybrid CFD--DEM solver \textit{SediFoam} is developed based on two state-of-the-art open-source
codes in their respective fields, i.e., a CFD platform OpenFOAM (Open Field Operation and
Manipulation) developed by OpenFOAM Foundation~\citep{openfoam} and a molecular dynamics simulator
LAMMPS (Large-scale Atomic/Molecular Massively Parallel Simulator) developed at the Sandia National
Laboratories~\citep{lammps}. The LAMMPS--OpenFOAM interface is implemented for the communication of
the two solvers. The code is publicly available at
{\color{blue}\url{https://github.com/xiaoh/sediFoam}} under GPL license. Detailed implementation is
discussed in~\cite{sun16sedi}.

The fluid equations in~(\ref{eq:NS}) are solved in OpenFOAM with the finite volume method
\citep{jasak96ea}. PISO (Pressure Implicit Splitting Operation) algorithm is used to prevent
velocity--pressure decoupling~\citep{issa86so}. A second-order central scheme is used for the
spatial discretization of convection terms and diffusion terms. Time integrations are performed with
a second-order implicit scheme. An averaging algorithm based on diffusion is implemented to obtain
smooth $\varepsilon_s$, $\mathbf{U}_s$ and $\mathbf{F}^{fp}$ fields from discrete sediment
particles~\citep{sun15db1, sun15db2}. To resolve the collision between the sediment particles, the
contact force between sediment particles is computed based on the non-linear contact theory
of~\cite{hertz1882udb}. The time step to resolve the particle collision is less than the Rayleigh
time of smallest particle and characteristic time of the fastest moving
particle~\citep{kremmer01amf,ji06eff}. The cohesive force model mentioned in Section~\ref{sec:dem}
is validated via fluidized bed simulations~\citep{gupta15vv}.

\section{Results}
\label{sec:sedi-tests}

Based on the above-mentioned approach, numerical simulations are performed to study the settling
process of cohesive silt particles. The first objective of the numerical simulations is to
demonstrate that CFD--DEM is capable of predicting the cohesive behaviors in the silt settling
process. Therefore, the cohesive behaviors obtained in the present simulations are compared with
available data in the literature, including (1) the decrease of the structural densities in the silt
beds, (2) the decrease in the separation between the characteristic lines due to flocculation, (3)
the variation of settling velocity in the settling process, and (4) the variation of the ratio
between van der Waals force and structural density. Another objective is to investigate the
influence of Bond number and particle polydispersity on silt sedimentation. Therefore, the variation
of structural densities on the silt beds due to the variation of Bond number and particle
polydispersity is detailed.  The setup of the numerical tests is based on previous experimental and
numerical studies~\citep{dong09dem,van13hs,te15hindered}.

\begin{figure}[htbp]
  \centering
  \includegraphics[width=0.45\textwidth]{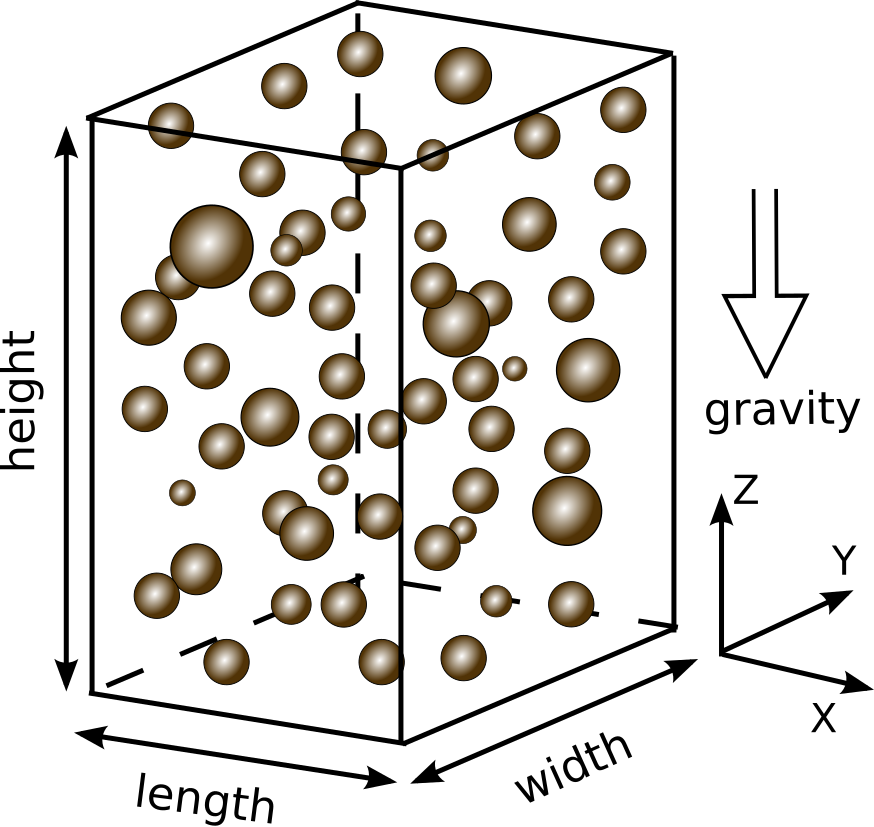}
  \caption{Layout of the numerical simulations of silt settling. The fluid is quiescent
    initially. The silt particles are settling from random locations and the initial solid volume
    fraction is uniform. Note that the particle size is not drawn to scale.}
  \label{fig:layout}
\end{figure}

The parameters used in the numerical simulations are presented in Table~\ref{tab:param-all},
including the dimensions of the domain, the mesh resolutions, and the properties of fluid and
particle phases. The geometry of the domain is shown in Fig.~\ref{fig:layout}.  Although the
dimensions of the computational domain are much smaller than those in the experimental study, the
size of the computational domain is adequate for CFD--DEM simulations of
sedimentation~\citep{dong06role,dong09dem,zhao14ig}. The $x$-, $y$- and $z$- coordinates are aligned
with the length, width, and height directions, respectively. The CFD mesh in all directions is
uniform in size. The boundary conditions for the pressure and the velocity fields are periodic in
both $x$- and $y$-directions. In $z$-direction, no-slip wall condition is applied on the bottom,
whereas pressure outlet boundary condition is applied on the top. The fluid is quiescent initially.
The silt particles are settling from random locations and the initial solid volume fraction is
uniform. The restitution coefficient of the sediment particles used in the simulations is very small
($e = 0.01$) because the contact among the sediment particles is viscously
damped~\citep{schmeeckle2014ns,kidanemariam14dns}. The coefficient of friction between the particles
is 0.4 according to~\cite{kidanemariam14dns}. It should be noted that both the restitution
coefficient and the coefficient of friction do not significantly influence the integral quantities
in sediment transport according to literature~\citep{drake01dp}. The fraction of silt by material
weight is approximated by using log-normal distribution~\citep{vanoni06se,garcia08se}. The process
to generate silt particles of random sizes based on the distribution of material weight is detailed
in the Appendix.

We simulated 14 different combinations of initial solid volume fraction $\varepsilon_{s,0}$,
particle polydispersity $d_{90}/d_{10}$, and Bond number Bo. The summary of different combinations
(Mono01 to Mono05, Poly01 to Poly09) is shown in Table~\ref{tab:param-run}.  The influences of Bond
number on the settling of mono- and poly-dispersed silt particles are investigated by using Mono01
to Mono05 and Poly01 to Poly05, respectively. The present simulation covers the range of Bond number
between 0 and 4. In addition to the influence of Bond number, the different cohesive behaviors of
mono- and poly-dispersed particles are compared. The influence of the initial solid volume fraction
on the settling process is investigated by comparing the results obtained in Poly04, and from Poly06
to Poly09. The initial solid volume fraction $\varepsilon_{s,0}$ ranges from 0.02 to 0.16 in the
simulations.  It is noted that the Hamaker coefficient is constant for particles of different sizes
in each simulation. This is because that the Hamaker coefficient does not change significantly for
the same material~\citep{israelachvili11is}. For poly-dispersed silt particle, the Bond number is
calculated based on Eq.~(\ref{eq:bond}) using the maximum cohesive force and the weight of mass
median particle diameter $d_{50}$.

\begin{table}[htbp]
 \caption{Parameters of the numerical simulations.}
 \begin{center}
 \begin{tabular}{lcccc}
   \hline       parameter & value range\\
   \hline
   domain dimensions                            &   \\
   \qquad length, width, height ($L_x/d_{50}, L_y/d_{50}, L_z/d_{50}$)   
   & \multicolumn{1}{ c }{$12\times12\times36$} \\
   mesh resolutions                             &   \\
   \qquad length, width, height ($N_x, N_y, N_z$)   
   & \multicolumn{1}{ c }{$12\times12\times36$} \\
   particle properties & \\
   \qquad initial solid volume fraction $\varepsilon_{s,0}$~[--] & \multicolumn{1}{ c }{[0.01, 0.16]}\\
   \qquad mass median particle diameter $d_{50}$~[$\mu$m]   & \multicolumn{1}{ c }{60}\\
   \qquad density $\rho_s$~[$\times 10^3~\mathrm{kg/m^3}$]  & \multicolumn{1}{ c }{2.7} \\
   \qquad Young's modulus~[MPa]                 & \multicolumn{1}{ c }{10} \\
   \qquad normal restitution coefficient~[--]    & \multicolumn{1}{ c }{0.1} \\
   \qquad coefficient of friction~[--]           & \multicolumn{1}{ c }{0.4} \\
   \qquad Hamaker coefficient[$\times 10^{-20}$]    & \multicolumn{1}{ c }{[0.0, 0.48]} \\
   \qquad Bond number~[--]                       & \multicolumn{1}{ c }{[0, 4]} \\
   fluid conditions & \\
   \qquad viscosity $\mu$~[$\times 10^{-3}~\mathrm{m^2/s}$] & \multicolumn{1}{ c }{1.0} \\
   \qquad density $\rho_f$~[$\times 10^3~\mathrm{kg/m^3}$]  & \multicolumn{1}{ c }{1.0} \\
   \hline
  \end{tabular}
 \end{center}
 \label{tab:param-all}
\end{table}

\begin{table}[!htbp]
 \caption{Conditions for numerical simulation tests.}
 \begin{center}
 \begin{tabular}{lccc}
   \hline
   \multirow{2}{*}{Case}    & initial volume fraction   & particle polydispersity   & Bond number \\
                            & ($\varepsilon_{s,0}$)     & ($d_{90}/d_{10}$)         & (Bo) \\
   \hline
   Mono01 & 0.16  & 1.0 & 0.0 \\
   Mono02 & 0.16  & 1.0 & 0.5 \\
   Mono03 & 0.16  & 1.0 & 1.0 \\
   Mono04 & 0.16  & 1.0 & 2.0 \\
   Mono05 & 0.16  & 1.0 & 4.0 \\
   Poly01 & 0.16  & 4.0 & 0.0 \\
   Poly02 & 0.16  & 4.0 & 0.5 \\
   Poly03 & 0.16  & 4.0 & 1.0 \\
   Poly04 & 0.16  & 4.0 & 2.0 \\
   Poly05 & 0.16  & 4.0 & 4.0 \\
   Poly06 & 0.01  & 4.0 & 2.0 \\
   Poly07 & 0.02  & 4.0 & 2.0 \\
   Poly08 & 0.04  & 4.0 & 2.0 \\
   Poly09 & 0.08  & 4.0 & 2.0 \\
   \hline
  \end{tabular}
 \end{center}
 \label{tab:param-run}
\end{table}

\subsection{Structural Density}
\label{sec:sedi-density}

In the sedimentation process of fluid--particle mixture, the height of the sediment bed increases.
Due to the formation of flocs by silt particle, the structural densities (solid volume fraction in
unity) of silt beds are smaller than those of sand beds~\citep{te15hindered}. In the present study,
the influences of inter-particle cohesion and particle polydispersity on the structural densities of
sediment beds are investigated separately. 

\begin{figure}[htbp]
  \centering
  \subfloat[non-cohesive, mono-disperse (Mono01)]{
  \includegraphics[width=0.45\textwidth]{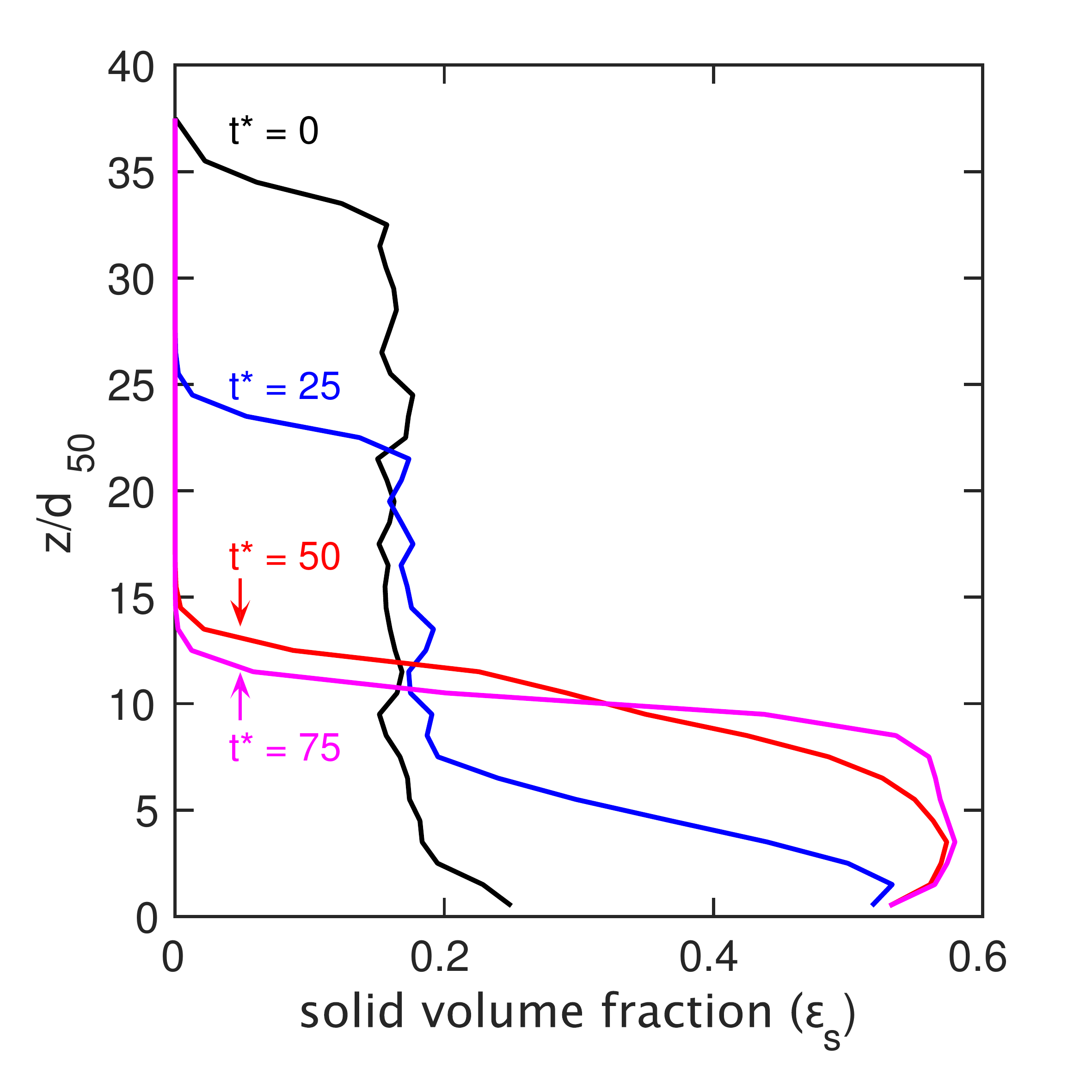}
  }
  \subfloat[cohesive, mono-disperse (Mono04)]{
  \includegraphics[width=0.45\textwidth]{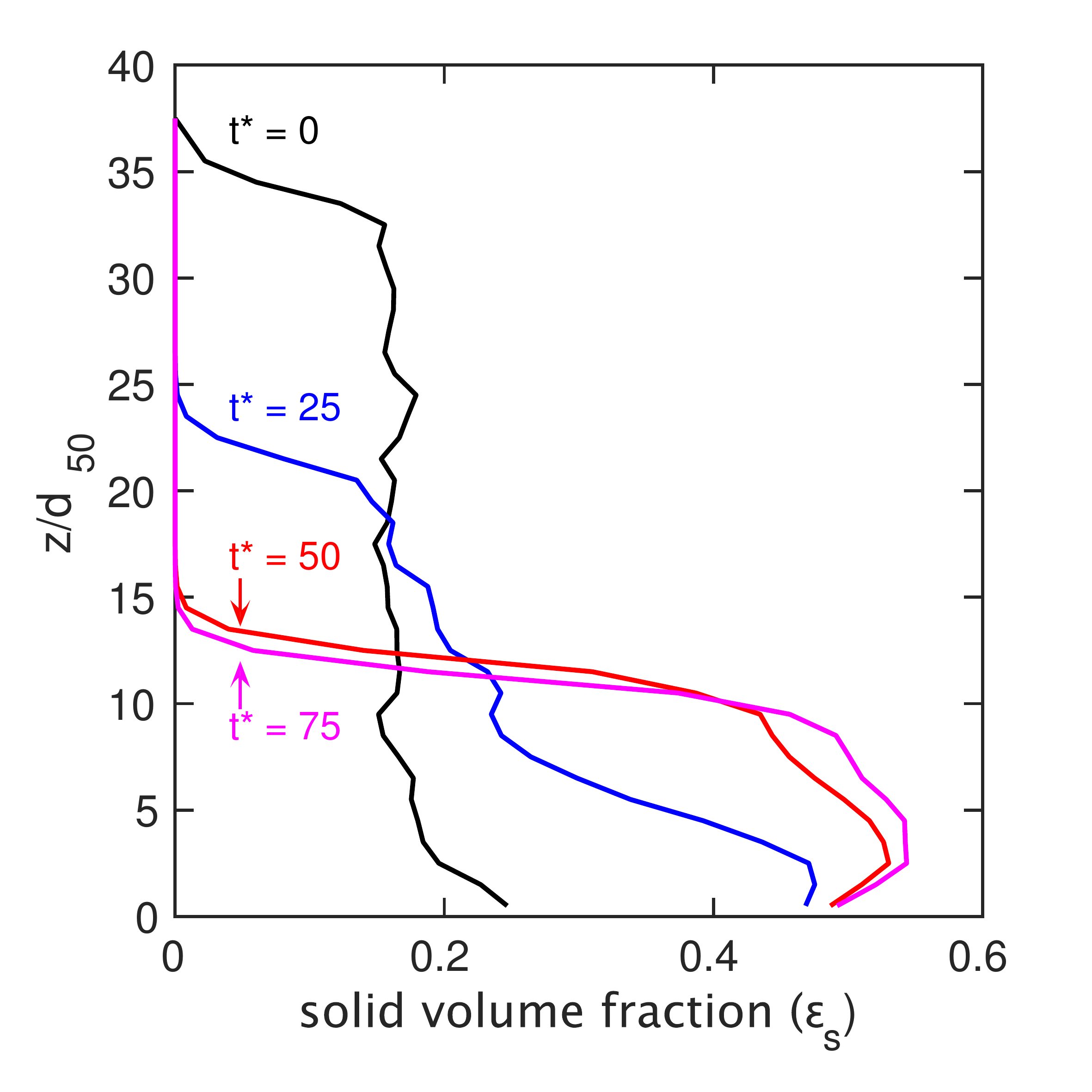}
  }
  \vspace{0.1in}
  \subfloat[non-cohesive, poly-disperse (Poly01)]{
  \includegraphics[width=0.45\textwidth]{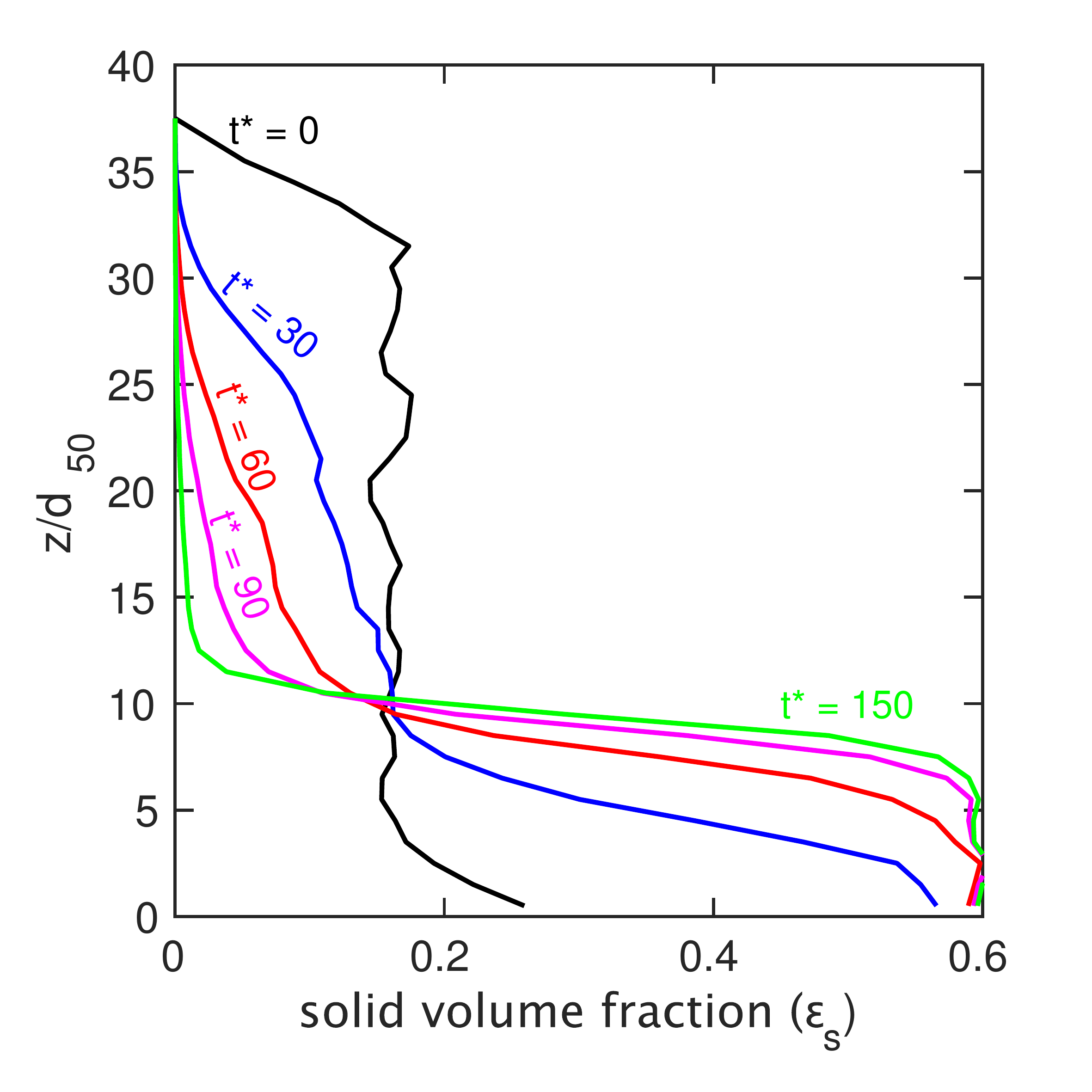}
  }
  \subfloat[cohesive, poly-disperse (Poly04)]{
  \includegraphics[width=0.45\textwidth]{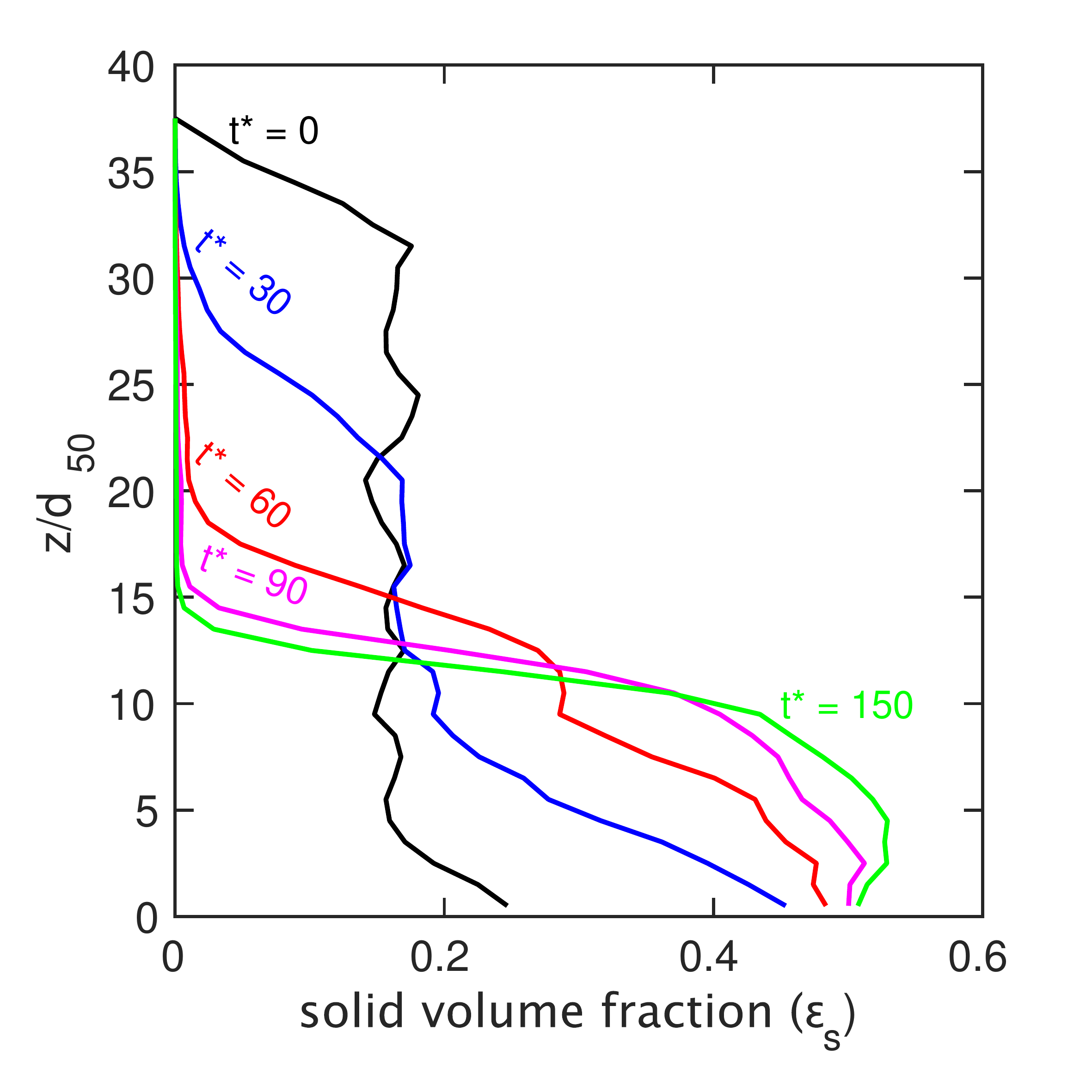}
  }
  \caption{Comparison of the solid volume fraction profiles of silt at different combinations of
    conditions.  The mass median particle diameter $d_{50} = 60$~$\mathrm{\mu m}$ for all test
    cases.  The initial solid volume fraction $\varepsilon_{s,0} = 0.16$. The polydispersity
    ($d_{90}/d_{10}$) is 1 and 4 for mono- and poly-dispersed particles, respectively. The Bond
    number (Bo) is 0 and 2 for non-cohesive and cohesive particles, respectively.}
  \label{fig:sedi-interface-all}
\end{figure}

The evolution of solid volume fraction profiles during the sedimentation process for silt particles
is presented in Fig.~\ref{fig:sedi-interface-all}. This aims to demonstrate that CFD--DEM is capable
of modeling the variation of the interfaces of the fluid--particle mixture. It can be seen in
Fig.~\ref{fig:sedi-interface-all} that the downward moving (supernatant/suspension) and upward
moving (suspension/sediment bed) interfaces are observed in the sediment test. This is consistent
with both experimental measurements and Kynch's theory on sedimentation~\citep{kynch52theory,
kranenburg92hs}. It can be seen in the figure that the stacking arrangement of the sediment
particles is influenced by wall boundary, and thus the solid volume fraction is smaller in the
near-wall regions. However, this does not influence the prediction of structural densities since the
thickness of the sediment bed is large enough~\citep{benenati62void}. The non-dimensional time $t^*
=  t w_{s,50}/d_{50}$ is used, where $w_{s,50}$ is the clear water terminal velocity of the
particles of mass median diameter. At $t^* = 0$, the initial solid volume fraction in the domain is
$\varepsilon_{s,0} = 0.16$. 

The solid volume fraction profiles of mono-dispersed particles are shown in
Figs.~\ref{fig:sedi-interface-all}(a)~and~\ref{fig:sedi-interface-all}(b). It can be seen 
that the shape of the profiles and the height of the downward moving interface at different
snapshots have no major difference between cohesive and non-cohesive particle. However, the
structural density in the sediment bed of cohesive particle is smaller than that of
non-cohesive particle after the particles are settled out of the suspension ($t^* = 75$). This
observation is similar to previous findings in cake sedimentation of glass beads~\citep{dong09dem},
indicating the decrease of structural density is due to particle cohesion even though the particles
are mono-dispersed.  Figures~\ref{fig:sedi-interface-all}(c)~and~\ref{fig:sedi-interface-all}(d)
show the solid volume fraction profiles of poly-dispersed particles. Due to the presence of smaller
particles in poly-dispersed silt, the sedimentation of poly-dispersed silt takes longer ($t^* =
150$) compared with that of mono-dispersed silt ($t^* = 75$). It can be seen in
Fig.~\ref{fig:sedi-interface-all}(c) that there is no downward moving interface in the sedimentation
process of poly-dispersed particles compared with mono-dispersed particles in
Fig.~\ref{fig:sedi-interface-all}(a). Instead, a gradient of the solid volume fraction in the
suspension can be observed. This gradient is attributed to the separation of poly-dispersed silt in
the settling process. The larger silt particles fall more rapidly to form the silt bed, whereas
small silt particles fall slowly in the suspension. In the upper part of suspension, there are only
very small particles, and the solid volume fraction is small; in the lower part, large and small
silt particles coexist, and the solid volume fraction is close to the initial value
$\varepsilon_{s,0}$. When the particle cohesion is considered in the poly-dispersed silt, the
smaller particles are attached to larger ones and fall faster. Therefore, the separation of
poly-dispersed silt is less significant and the gradient of the solid volume fraction in the
suspension is smaller, as is shown in Fig.~\ref{fig:sedi-interface-all}(d). It can be seen that the
range of the solid volume fraction in the sediment bed of cohesive poly-dispersed silt is [0.48,
0.53] at Bond number Bo = 2. This range is qualitatively consistent with the result [0.46, 0.55] in
experimental measurement~\citep{te15hindered}.  When the Bond number = 2, the Hamaker coefficient in
the simulation equals to $0.24\times10^{-20}$, which of the same order of magnitude $O(10^{-20})$ as
in the literature~\citep{israelachvili11is}.

\begin{figure}[htbp]
  \centering
  \includegraphics[width=0.55\textwidth]{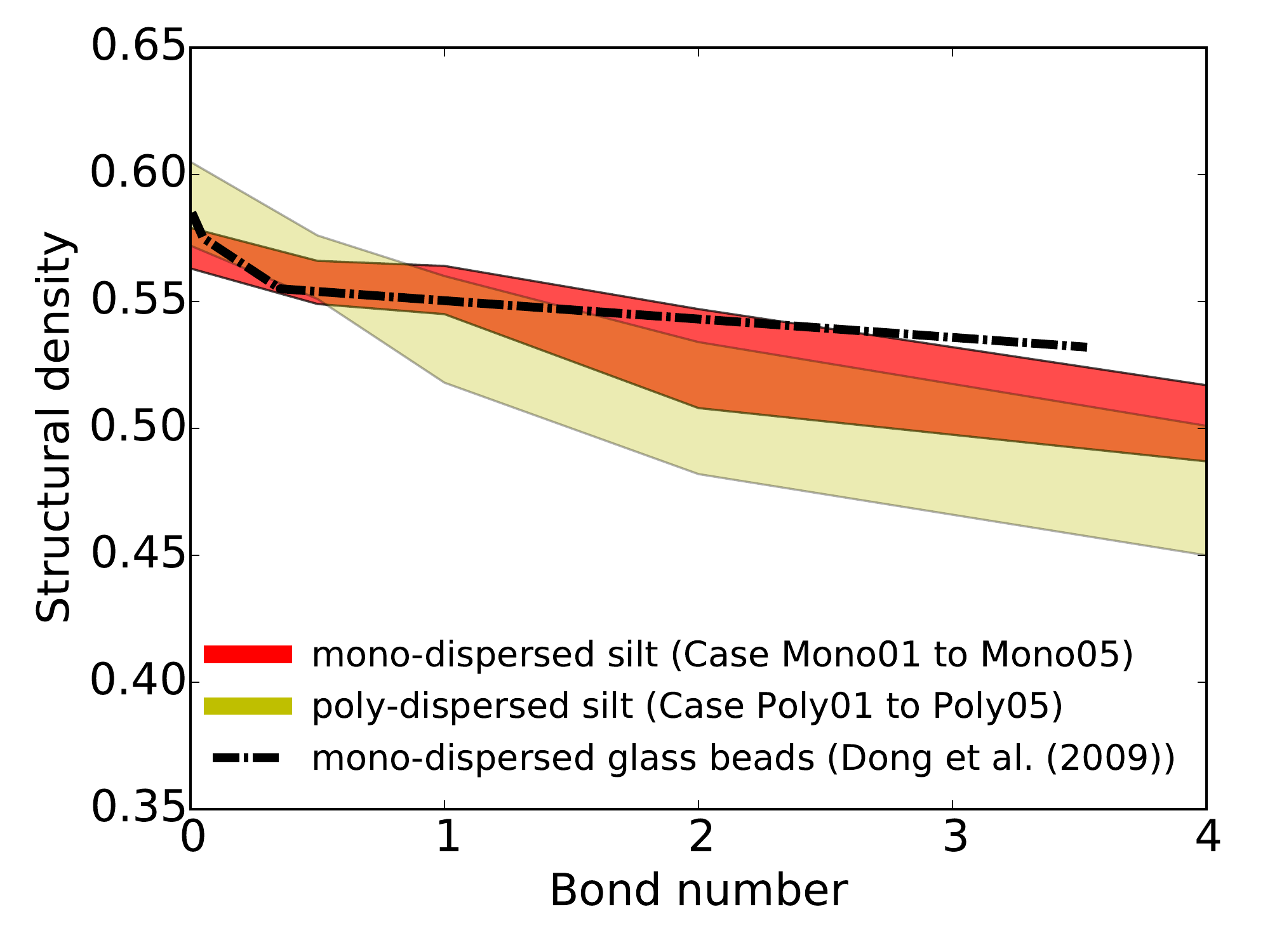}
  \caption{Comparison of the structural densities of silt particle at different Bond numbers with
    the CFD--DEM simulation results using cohesive glass beads in the literature~\citep{dong09dem}.
    The shade indicates the range of the structural densities obtained in present simulations.}
  \label{fig:sedi-bond}
\end{figure}

The structural densities of the sediment beds of mono- and poly-dispersed silt obtained in the
present simulations are plotted as a function of the Bond number in Fig.~\ref{fig:sedi-bond}. The
range of the structural densities is used to demonstrate its variation within the sediment beds.
This aims to illustrate the influence of both Bond number and particle polydispersity on the
structural densities of the sediment beds. It can be seen that the variations of the structural
densities at different Bond numbers in the present simulations are consistent with those obtained in
previous CFD--DEM simulations of much larger particles~\citep{dong09dem}. This indicates that the
influence of the Bond number on the decrease of the structural densities is similar for different
particle sizes. It can be seen that the structural densities of poly-dispersed silt when Bond number
Bo $< 1$ are larger than that of mono-dispersed silt.  This is because smaller particles can fill in
the spaces between larger particles and have larger structural densities when the cohesive force is
negligible. However, at Bond number Bo $\ge 1$, the structural densities of poly-dispersed silt are
much smaller. This is attributed to the fact that the number of particles smaller than $d_{50}$ is
very large (see Fig.~\ref{fig:particleSize}(b)), and the influence of cohesive force is more
significant on smaller particles. The flocs of smaller silt particles are more likely to form, and
the cohesive force limited the relative motion of silt particles to  prevent the formation of more
close-packed structures. Moreover, the variation of structural densities is larger for
poly-dispersed silt. This is due to the variation of particle size in poly-dispersed silt beds.

In summary, the structural densities of both mono- and poly-dispersed silt at various Bond numbers
are presented. The results obtained in the present studies are qualitatively consistent with those
in the literature. It can be seen that the cohesion from silt particle itself can contribute to the
decrease of structural densities. Moreover, the presence of smaller particles in poly-dispersed silt
increases the reduction of structural densities.

\subsection{Characteristic Lines}
\label{sec:sedi-line}
The characteristic lines of equal solid volume fraction (isolutes) are constructed from solid volume
fraction profiles. The descending and ascending of isolutes indicate the segregation of particles of
different sizes. In this section, the characteristic lines obtained from CFD--DEM simulations are
detailed, which is to demonstrate the capability of the CFD--DEM approach in the prediction of
segregation of cohesive silt. 

\begin{figure}[htbp]
  \centering
  \includegraphics[width=0.8\textwidth]{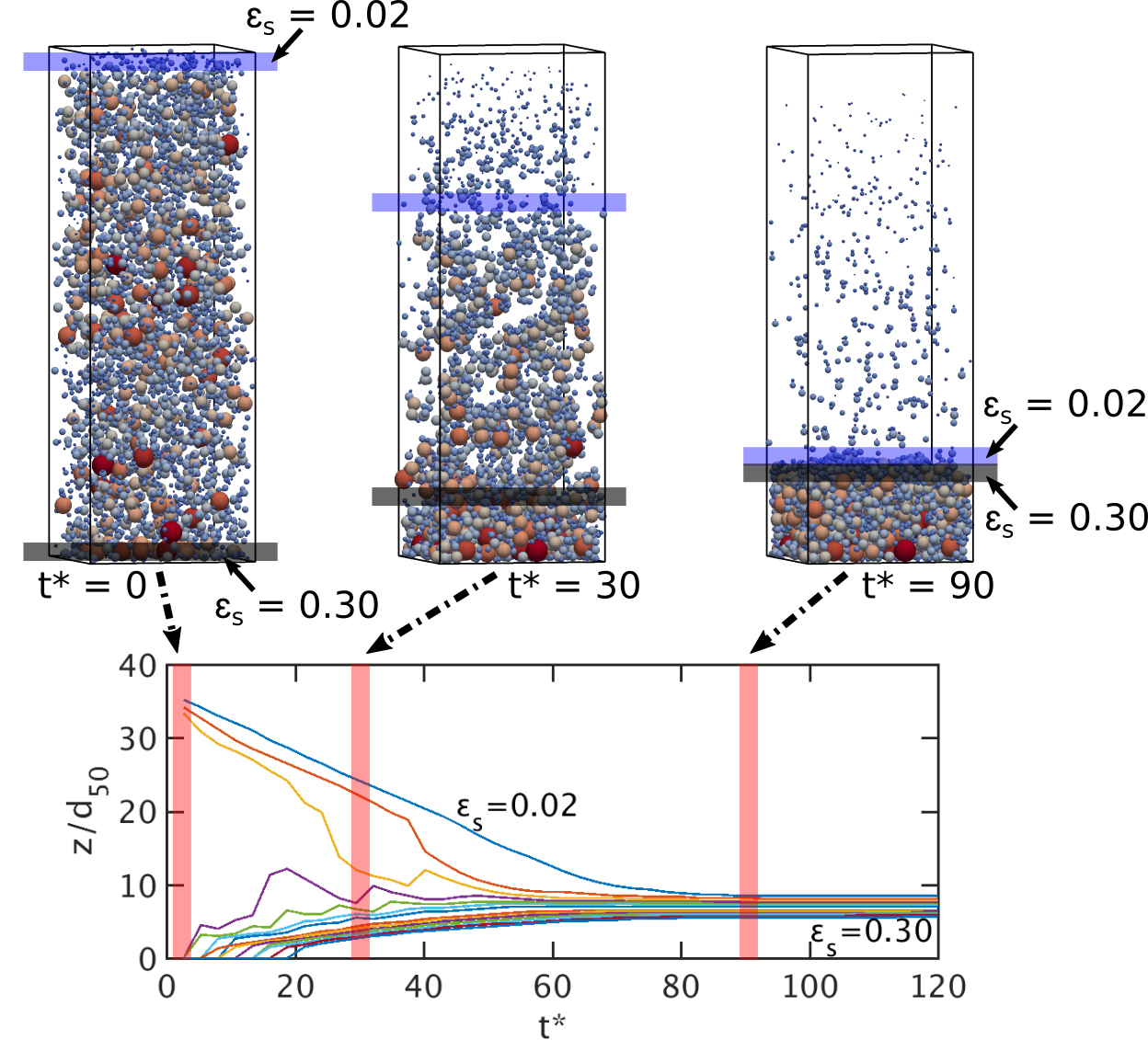}
  \caption{Evolution of the settling process of poly-dispersed silt (Poly09). Three representative
    snapshots in the settling process of silt particles at $t^* = $ 0, 30, and 90 are presented in
    the upper part. The lower part shows the characteristic lines (isolutes of sediment solid volume
    fraction) in the ($z,t$) plane. The initial solid volume fraction $\varepsilon_{s,0} = 0.08$,
    the Bond number Bo = 2, and particle polydispersity $d_{90}/d_{10} = 4$. The isolutes are
    plotted for $\varepsilon_{s} \in [0,0.3]$ with the spacing $\Delta \varepsilon_s = 0.02$.}
  \label{fig:cha-evo}
\end{figure}

The evolution of characteristic lines in the ($z,t$) plane during the settling process for case
Poly09 is shown in Fig.~\ref{fig:cha-evo}. The descending and ascending of the isolutes of the
horizontally-averaged solid volume fraction fields are demonstrated. The descending of the isolutes
represents the settling of particles, whereas the ascending of the isolutes represents the accretion of
sediment beds. It can be seen in Fig.~\ref{fig:cha-evo} that the variation in slopes of the
descending characteristic lines (fan of isolutes) can be observed in the settling process of
poly-dispersed particles. This is due to the segregation of poly-dispersed silt, since the silt
particles are composed of different size fractions and have different settling velocities. The fan
of isolutes is also observed in the previous experimental study~\citep{te15hindered}. When the
sediment are settled out of the suspension, the isolutes indicate the surface of the sediment beds
and the characteristic lines become horizontal in the ($z,t$) plane.

\begin{figure}[htbp]
  \centering
  \subfloat[cohesive, $\varepsilon_{s,0} = 0.08$ (Poly09)]{
  \includegraphics[width=0.45\textwidth]{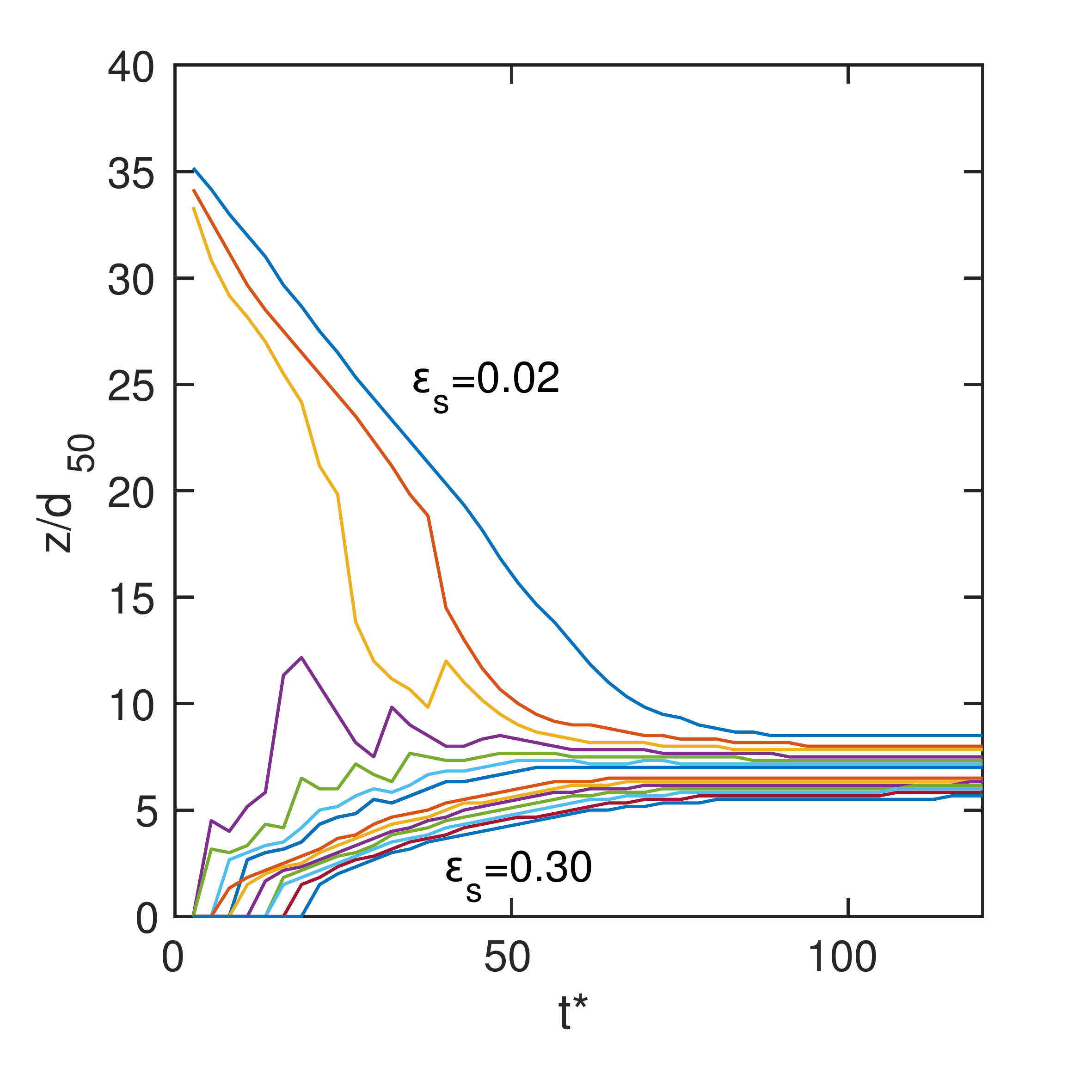}
  }
  \subfloat[cohesive, $\varepsilon_{s,0} = 0.16$ (Poly04)]{
  \includegraphics[width=0.45\textwidth]{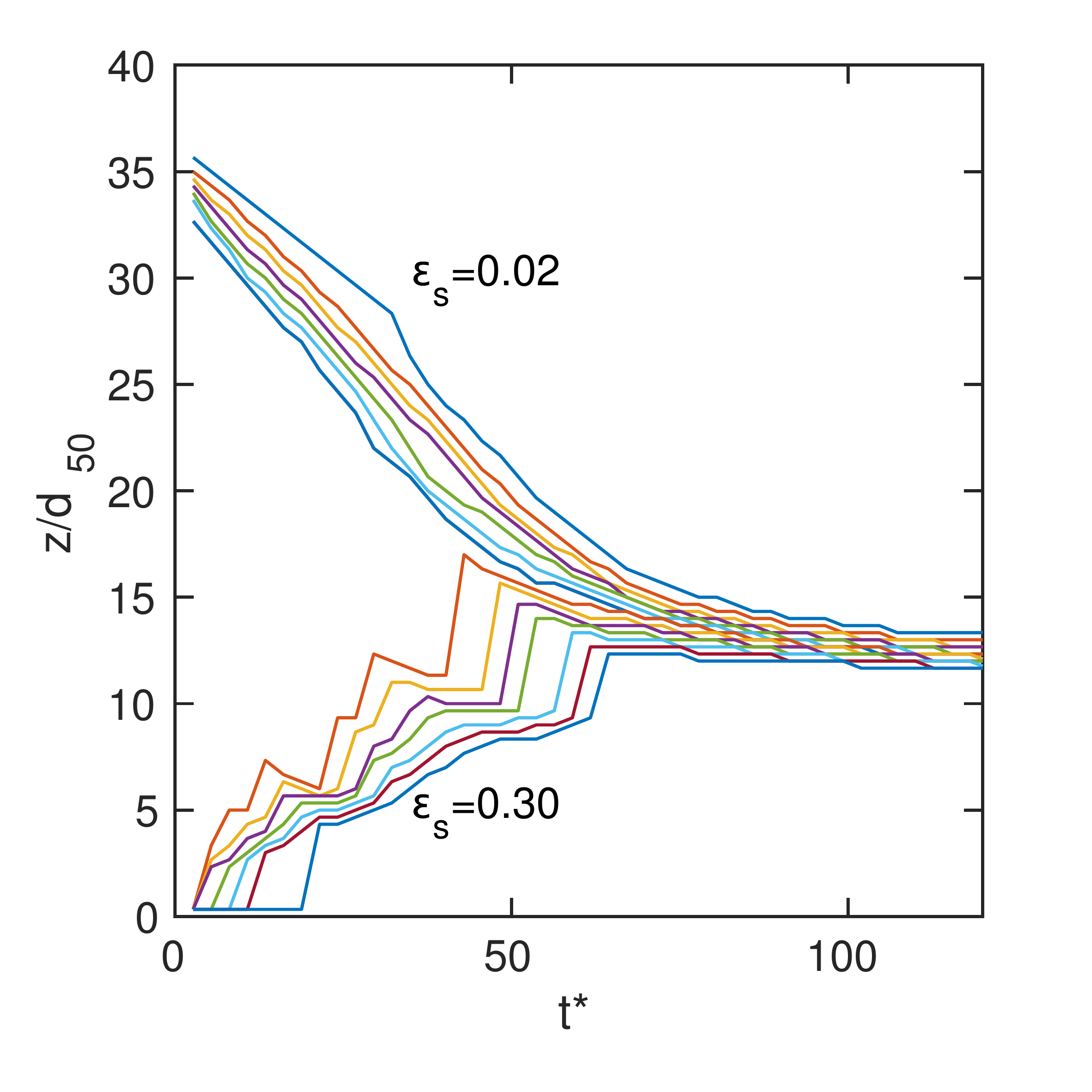}
  }
  \vspace{0.1in}
  \subfloat[non-cohesive, $\varepsilon_{s,0} = 0.16$ (Poly01)]{
  \includegraphics[width=0.45\textwidth]{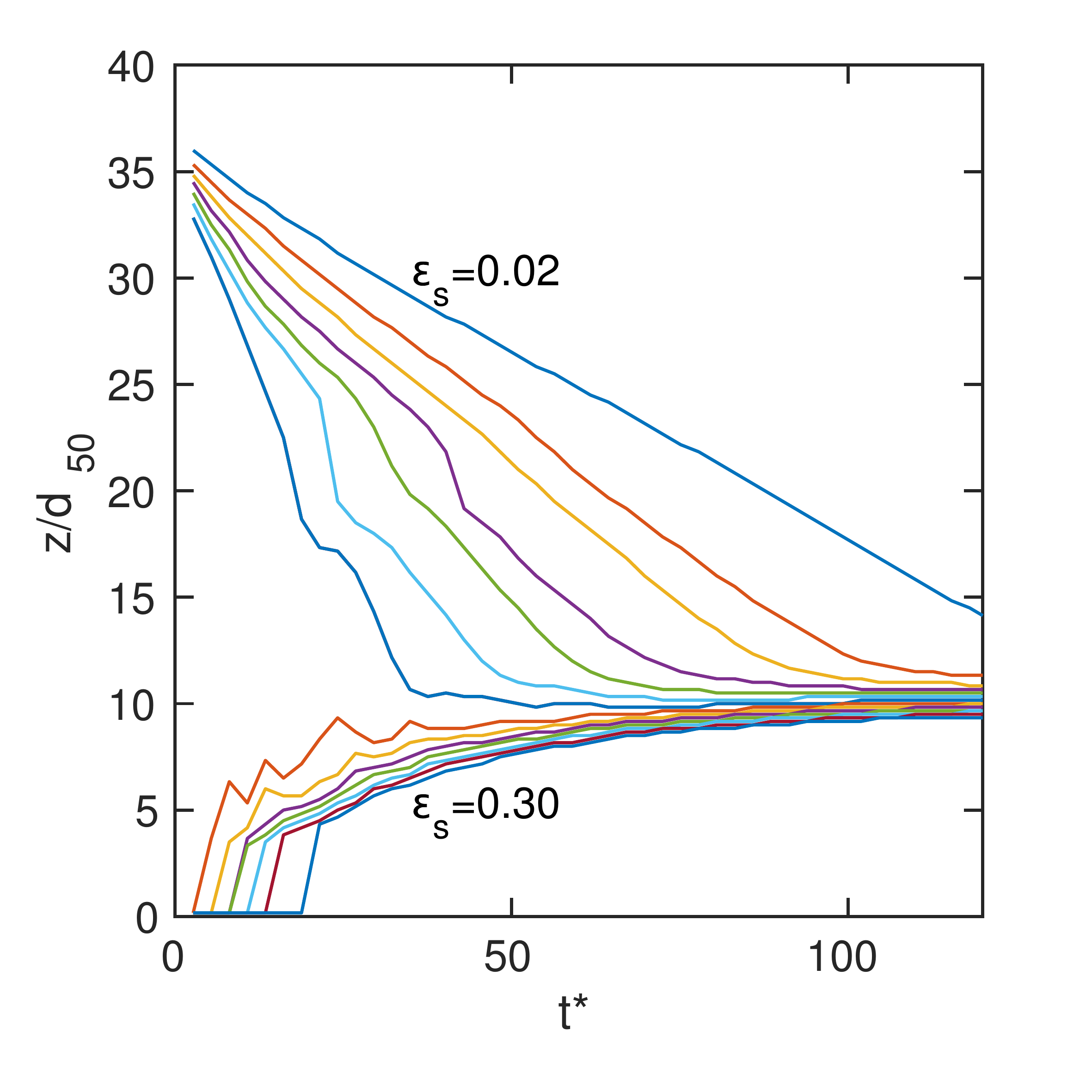}
  }
  \caption{Comparison of the characteristic lines (isolutes of sediment solid volume fraction) of
    poly-dispersed silt sedimentation tests. The isolutes are plotted for $\varepsilon_{s} \in
    [0,0.3]$ with the spacing $\Delta \varepsilon_s = 0.02$. The particle
    polydispersity $d_{90}/d_{10} = 4$.}
  \label{fig:cha-lines}
\end{figure}

The characteristic lines obtained from cohesive poly-dispersed silt at different initial solid
volume fractions $\varepsilon_{s,0}$ are shown in Figs.~\ref{fig:cha-lines}(a)
and~\ref{fig:cha-lines}(b). At both initial solid volume fractions, the fans of isolutes due to the
segregation of poly-dispersed silt are observed. However, when the initial solid volume fraction
increases from $\varepsilon_{s,0} = 0.08$~to~0.16, the segregation effect reduces significantly.
This is attributed to the fact that the increase of initial solid volume fraction can hinder the
segregation process~\citep{te15hindered}. The influence of the Bond number to characteristic lines
can be seen from the comparison between Figs.~\ref{fig:cha-lines}(b) and~\ref{fig:cha-lines}(c). The
rising isolutes are similar in both panels, but the shapes of the descending isolutes are different.
When the cohesive force is considered, the separation of the descending isolutes is negligible.
However, when the cohesive force is not considered, the separation is much more significant. This is
because when particle cohesion is considered, smaller particles are attached to larger ones and fall
much faster.  Consequently, the segregation due to the difference of particle settling velocity is
significantly reduced.  Compared with the experimental study, the present numerical simulation uses
a much smaller computational domain, and the characteristic lines obtained in numerical tests cannot
be directly compared to the experimental results. However, the qualitative features of the
characteristic lines observed in the experimental study~\citep{te15hindered} are still captured in
the numerical simulations.  

\subsection{Settling Velocity}
\label{sec:sedi-vel}
The settling velocity is another quantity of interest in the settling process. Since natural silt is
poly-dispersed and the settling velocities of different particle sizes differ significantly, the
settling velocity varies as the particle distribution of the suspended particles changes. To
describe the variation of settling velocity during the settling process of natural silt, the
combinations of settling velocity and solid volume fraction are used~\citep{te15hindered}. In the
present work, the settling velocities obtained by using CFD--DEM are presented to demonstrate the
capability of CFD--DEM in predicting the maximum settling velocity and the evolution of the settling
velocity during the settling process.

\begin{figure}[htbp]
  \centering
  \subfloat[]{
  \includegraphics[width=0.45\textwidth]{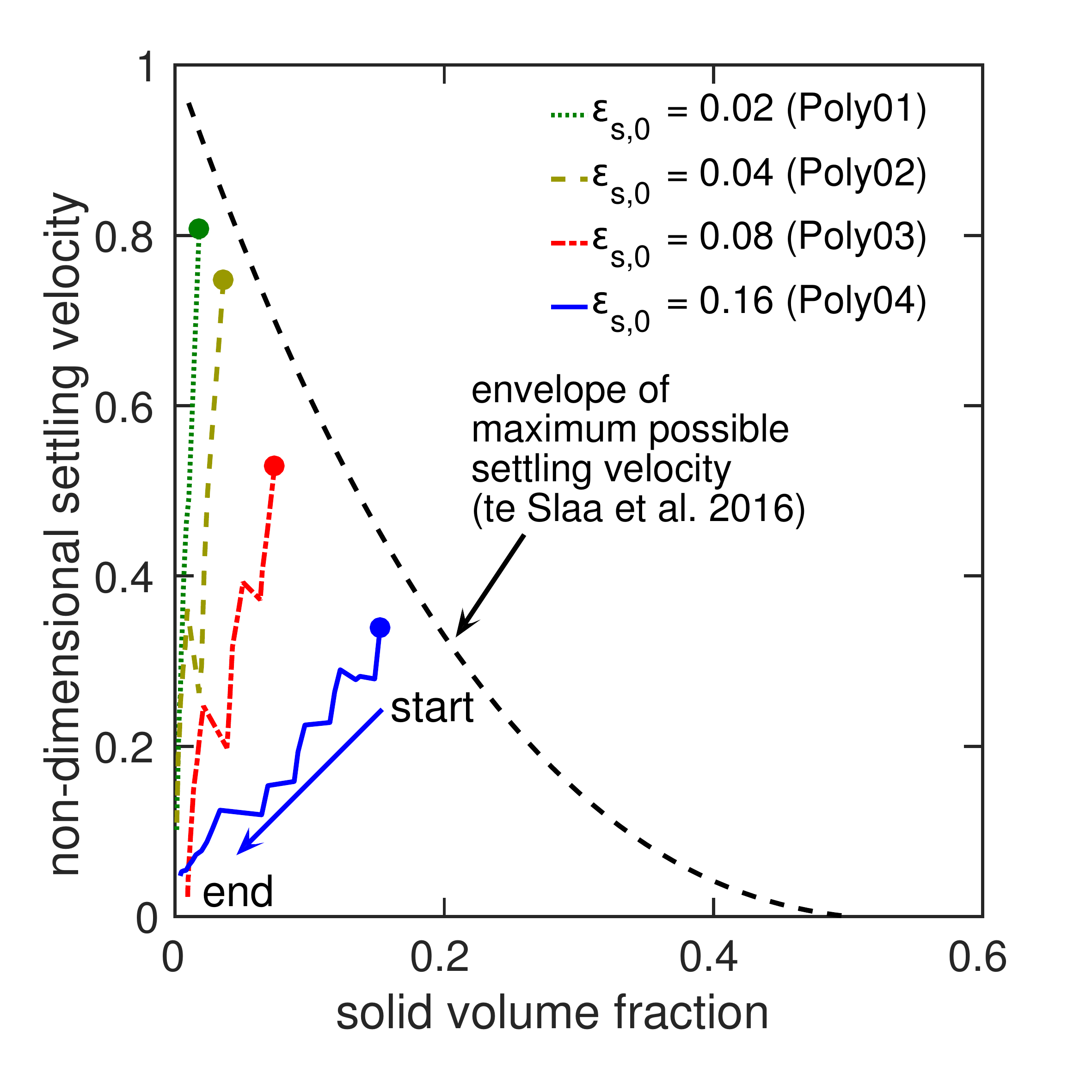}
  }
  \subfloat[]{
  \includegraphics[width=0.45\textwidth]{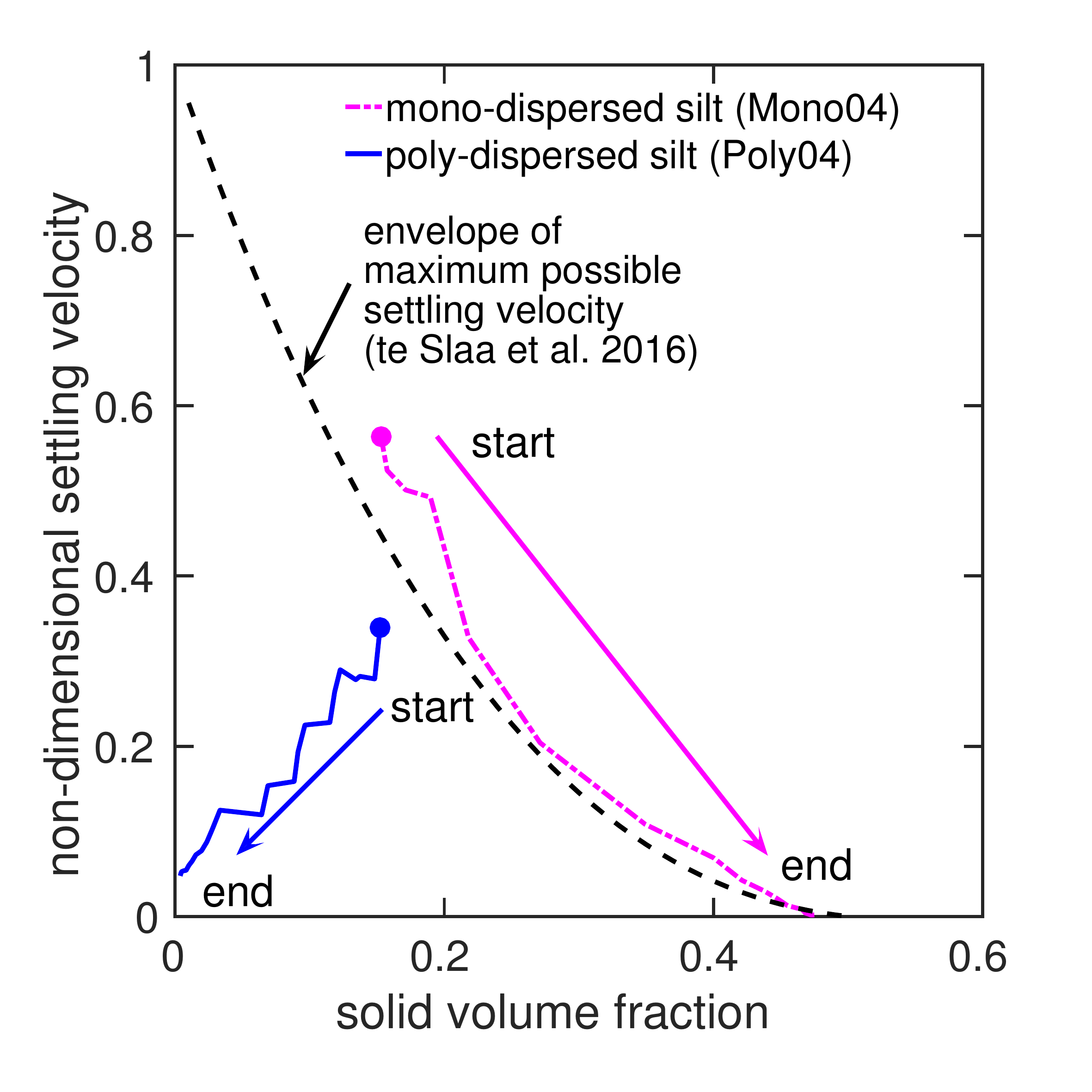}
  }
  \caption{Comparison of the silt particle terminal velocities with the envelope of maximum possible
    velocity based on the reference~\citep{te15hindered}. Panel (a) compares the particle terminal
    velocities obtained by using poly-dispersed silt particles at different initial solid volume
    fractions. Panel (b) shows the comparison between mono- and poly-dispersed silt particle at
    initial solid volume fraction $\varepsilon_{s,0} = 0.16$. The dots indicate the initial settling
    velocities in the settling process.}
  \label{fig:vel-envelope}
\end{figure}

The settling velocity is obtained by calculating the volume-averaged particle velocity of the moving
particles, excluding the stationary particles in the sediment bed. The settling velocities are
normalized by the clear water settling velocity ($w_{s,50}$). The evolution of settling
velocities and the solid volume fractions of silt particle suspension obtained in the present
simulations are shown in Fig.~\ref{fig:vel-envelope}. The settling velocities obtained at different
solid volume fractions according to the hindered settling function in the literature are plotted as
the enclosing envelope for all possible combinations~\citep{te15hindered}. The markers in
Fig.~\ref{fig:vel-envelope} indicate the initial settling velocities before the settling process.
The evolution of both solid volume fraction and settling velocity of the suspension are denoted
using arrows. 

The settling velocities of poly-dispersed silt at different initial solid volume fractions
$\varepsilon_{s,0}$ are shown in Fig.~\ref{fig:vel-envelope}(a). It can be seen in the results that
the initial settling velocities obtained at all particle solid volume fractions are close to the
envelope. This indicates the settling velocities predicted by using CFD--DEM are consistent with the
results obtained by using the empirical formula which considered the influence of the silt
flocculation and reverse flow~\citep{te15hindered}. Therefore, the CFD--DEM approach captures their
influence on the averaged settling velocities of silt particles. In addition, the settling
velocities predicted by present simulations of poly-dispersed silt are located below the velocity
envelope. This is consistent with both the experimental measurement and the sedimentation theory of
poly-dispersed particles~\citep{te13sp,te15hindered}. The comparison between mono- and
poly-dispersed silt is plotted in Fig.~\ref{fig:vel-envelope}(b).  It can be seen that the
evolution of solid volume fractions and settling velocities of mono-dispersed silt is different
from those of poly-dispersed silt. The solid volume fraction of the suspension of
mono-dispersed silt increases during the settling process (see
Fig.~\ref{fig:sedi-interface-all}(b)). It can be seen that the terminal velocities of the
mono-dispersed particles at different solid volume fractions are consistent with the maximum
possible settling velocities. This suggests that settling velocity of mono-dispersed silt predicted
by using CFD--DEM follows the empirical formula during the sedimentation and consolidation process
when the solid volume fractions of the suspension increases. 

\subsection{Force Ratio}
\label{sec:force-ratio}
When the settling process is terminated, the influence of fluid drag decreases and the
inter-particle contact increases to balance the gravity. The packing arrangement after settling can
be related to the ratio between total cohesive force and submerged particle
weight~\citep{yang00cs,dong09dem}.  Therefore, the force ratio is an important quantity of interest
to describe the influence of cohesive force on the packing arrangement. In this section, the force
ratios obtained in the present simulations are shown and compared with those in the literature. This
aims to demonstrate that correlations between the force ratio and the packing arrangement in the
silt bed are captured in present simulations.

\begin{figure}[htbp]
  \centering
  \includegraphics[width=0.65\textwidth]{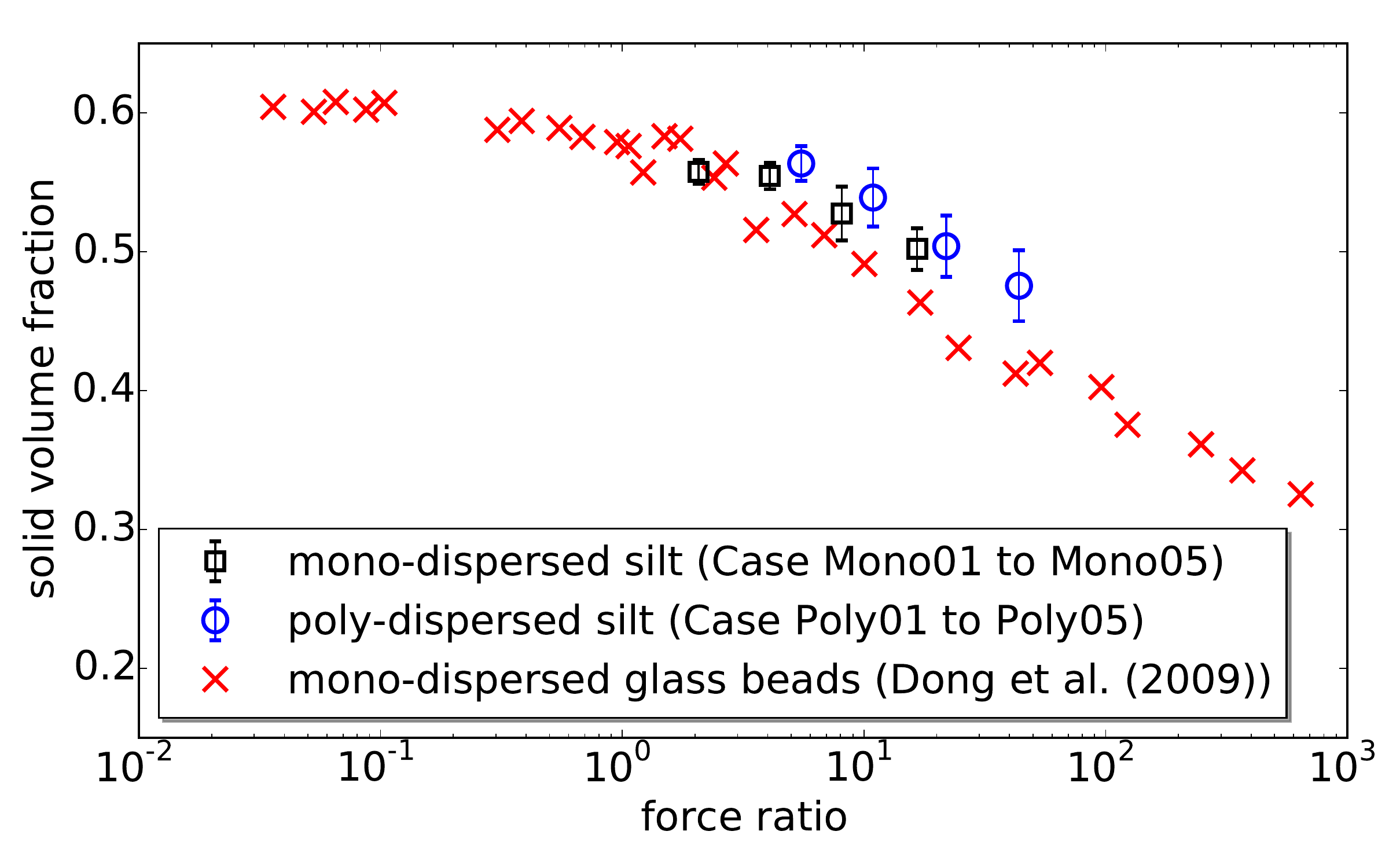}
  \caption{The solid volume fractions plotted as a function of the force ratio obtained in the
    present simulations. The results obtained in the literature by using CFD--DEM simulations of
    much larger cohesive glass beads are plotted for comparison~\citep{dong06role,dong09dem}. The
    error bar indicates the variation of solid volume fractions in the sediment beds obtained in the
    present simulations.}
  \label{fig:force-ratio}
\end{figure}

The ratio between the total cohesive force and submerged particle weight is defined
as~\citep{yang00cs,dong09dem}:
\begin{equation}
  \xi = |\sum_j \mathbf{f}^{vdw}_{ij}|/|\mathbf{F}_{submerged,i}|,
  \label{eq:force-ratio}
\end{equation}
where $\xi$ is the force ratio; $|\sum\limits_j \mathbf{f}^{vdw}_{ij}|$ and
$\mathbf{F}_{submerged,i} = (\rho_s - \rho_f)V_{p,i}\mathbf{g}$ are the total cohesive force and the
submerged weight on particle $i$, respectively.  The solid volume fractions of the silt bed and the
force ratios for both mono- and poly-dispersed particles obtained in the simulations are shown in
Fig.~\ref{fig:force-ratio}. When the cohesive force increases, the relative motion between silt
particles is limited and thus the packing arrangement in the sediment bed is loosen. Compared with
mono-dispersed particles, poly-dispersed particles have slightly larger force ratios. This is
because poly-dispersed particles are likely to have more contacts than mono-dispersed particles, and
thus the total cohesive force is larger. Moreover, it can be also seen that the relationship between
force ratios and solid volume fractions obtained in the present simulations is consistent with the
results in the literature~\citep{dong09dem}. It is noted that the diameter of silt ($d_{50} =
60$~$\mu$m) is smaller than the glass beads ($d_{50} = 250$~$\mu$m) in the literature. This suggests
the similar correlation relationship between force ratio and solid volume fraction is observed at
numerical tests using different particles sizes. 

\section{Discussion}

Currently, CFD--DEM are mostly constrained to numerical simulations in small domains (of the order
of centimeters) due to the high computational costs. However, engineering scales are much larger,
usually ranging from tens of meters to several kilometers. Therefore, macro-scale models have been
widely used to study the settling of cohesive sediments in engineering practices, but they need
ad-hoc closure models for the influence of flocculation. These ad-hoc models are not derived
directly from first principles and cause large discrepancies in the predictions of the cohesive
behavior of silt particles, particularly when used outside their applicable regimes.  CFD--DEM can
provide valuable physical insights of the cohesive particle flocculation, which can provide guidance
to the macro-scale modeling of the settling process using empirical models. To this end, we discuss
the cohesive force network and collision density functions obtained from the present simulations and
their implications for macro-scale modeling of cohesive silt sedimentation.

\begin{figure}[htbp]
  \centering
  \subfloat[$\varepsilon_{s,0} = 0.04$ (Poly08)]{
  \includegraphics[width=0.45\textwidth]{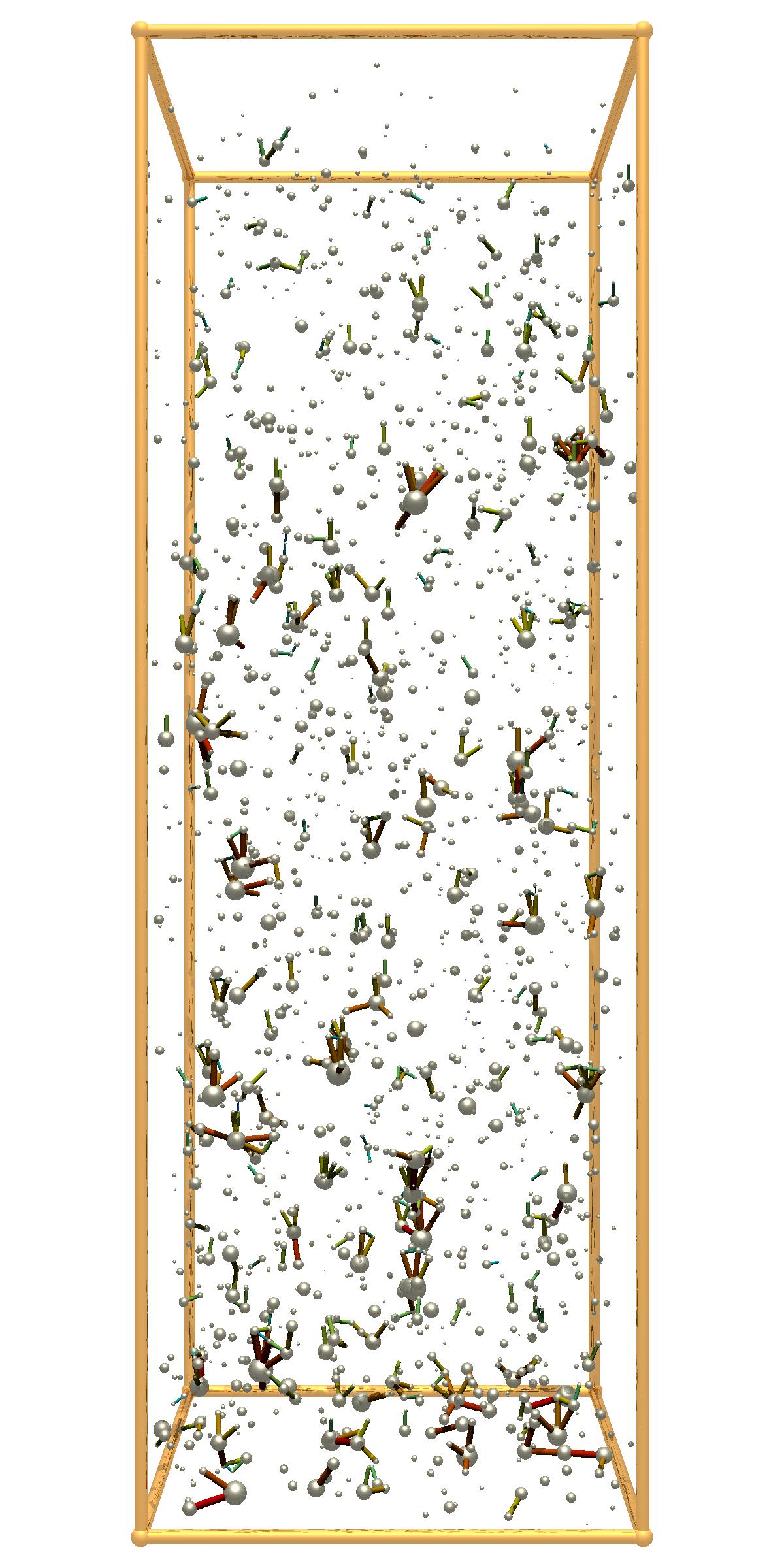}
  }
  \subfloat[$\varepsilon_{s,0} = 0.16$ (Poly04)]{
  \includegraphics[width=0.45\textwidth]{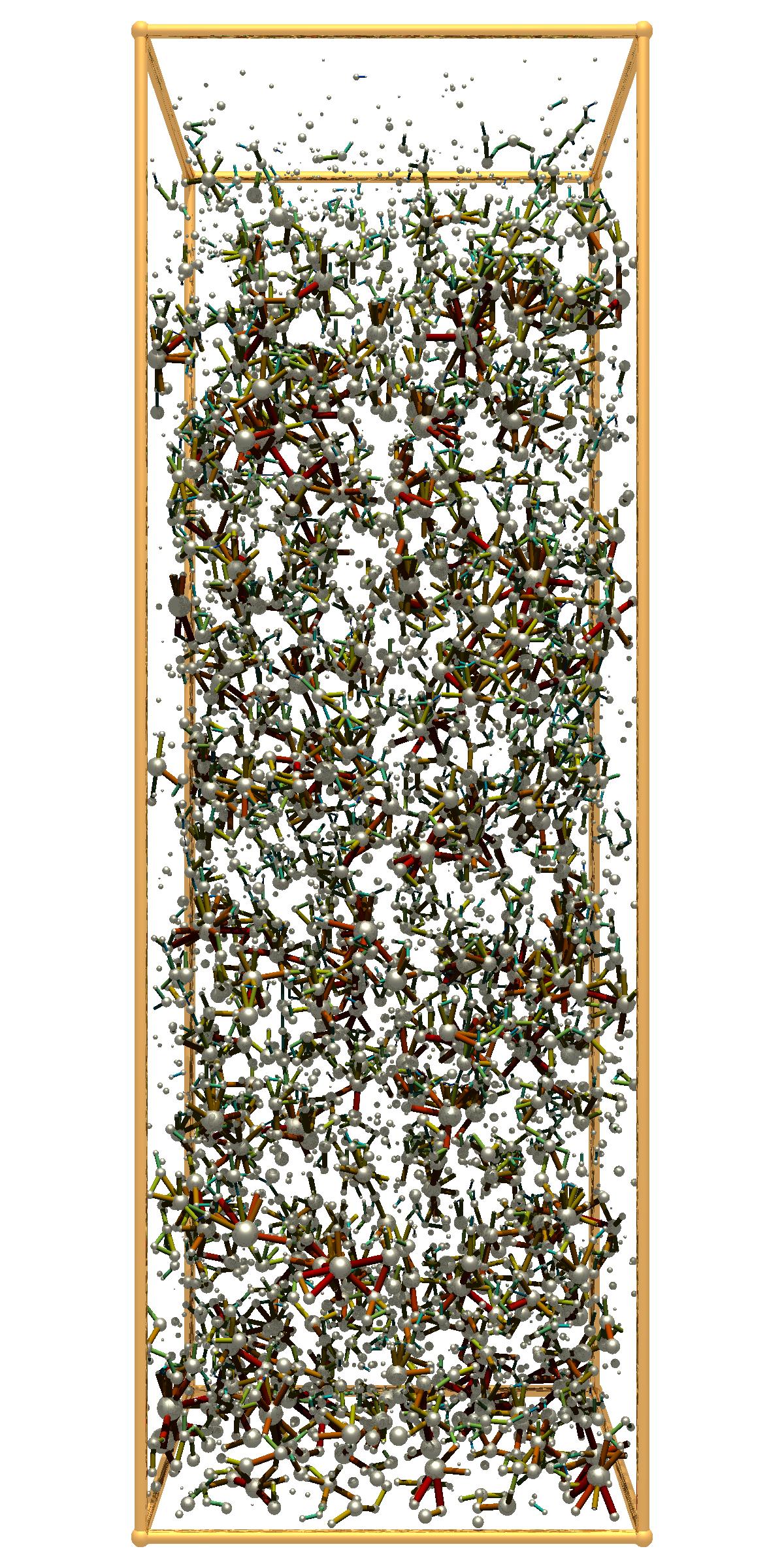}
  }
  \caption{The force network obtained at different initial solid volume fractions: (a)
    $\varepsilon_{s,0} = 0.04$; (b) $\varepsilon_{s,0} = 0.16$. The spheres indicate the size and
    positions of the silt particles, and the cylinders indicate the cohesive force between the
    particles. The simulations are performed at Bond number Bo = 2. Note that the particle sizes are
    scaled by a factor of 0.2.}
  \label{fig:cff-network}
\end{figure}

The cohesive force network between silt particles is shown in Fig.~\ref{fig:cff-network} to
demonstrate the propagation of the cohesive force. The locations of particles and cohesive force are
obtained at $t^* = 5.4$ before the sedimentation of larger particles. The particles are represented
by using spheres of different sizes. The cohesive forces between the particles are represented by
using cylinders, and the diameter of the cylinders indicates the magnitude of the cohesive force.
The cohesion between particles of different sizes can be observed in Fig.~\ref{fig:cff-network}.
This is caused by the differential settling of poly-dispersed silt. Because of the inter-particle
cohesion, the flocs of silt particle are formed at both initial solid volume fractions.  At lower
initial solid volume fraction (see Fig.~\ref{fig:cff-network}(a)), small and isolated flocs are
observed. On the other hand, at higher initial solid volume fraction (see
Fig.~\ref{fig:cff-network}(b)), the number of contacts between silt particles increases
significantly. The flocs are connected via van der Waals force and are no longer isolated as in the
case of low initial solid volume fractions. It can be also seen that the magnitude of the cohesive
force is larger between larger particles than that between smaller particles. However, the Bond
number between smaller particles is larger.

The collision frequency function has been used to model the rate of collision between cohesive
particles in macro-scale problems of silt sedimentation~\citep{van88agg,van94est}. This coefficient
is important in the prediction of the settling velocity when flocs are formed by cohesive particles.
In the modeling of flocculation based on binary collision theory, the collision frequency
function is defined as~\citep{van88agg}:
\begin{equation}
  K_{col} = \frac{\pi g}{72\rho_p\nu}\left( d_i + d_j \right)^2\left( \Delta\rho d_i^2 - \Delta\rho
  d_j^2\right),
  \label{eq:cff}
\end{equation}
where $\Delta \rho = \rho_s - \rho_f$ is the difference in density between fluid and particle.
However, the binary collision theory does not account for the influence of solid volume fraction on
the collision frequency.

\begin{figure}[htbp]
  \centering
  \subfloat[initial solid volume fraction $\varepsilon_{s,0} = 0.04$ (Poly08)]{
  \includegraphics[width=0.45\textwidth]{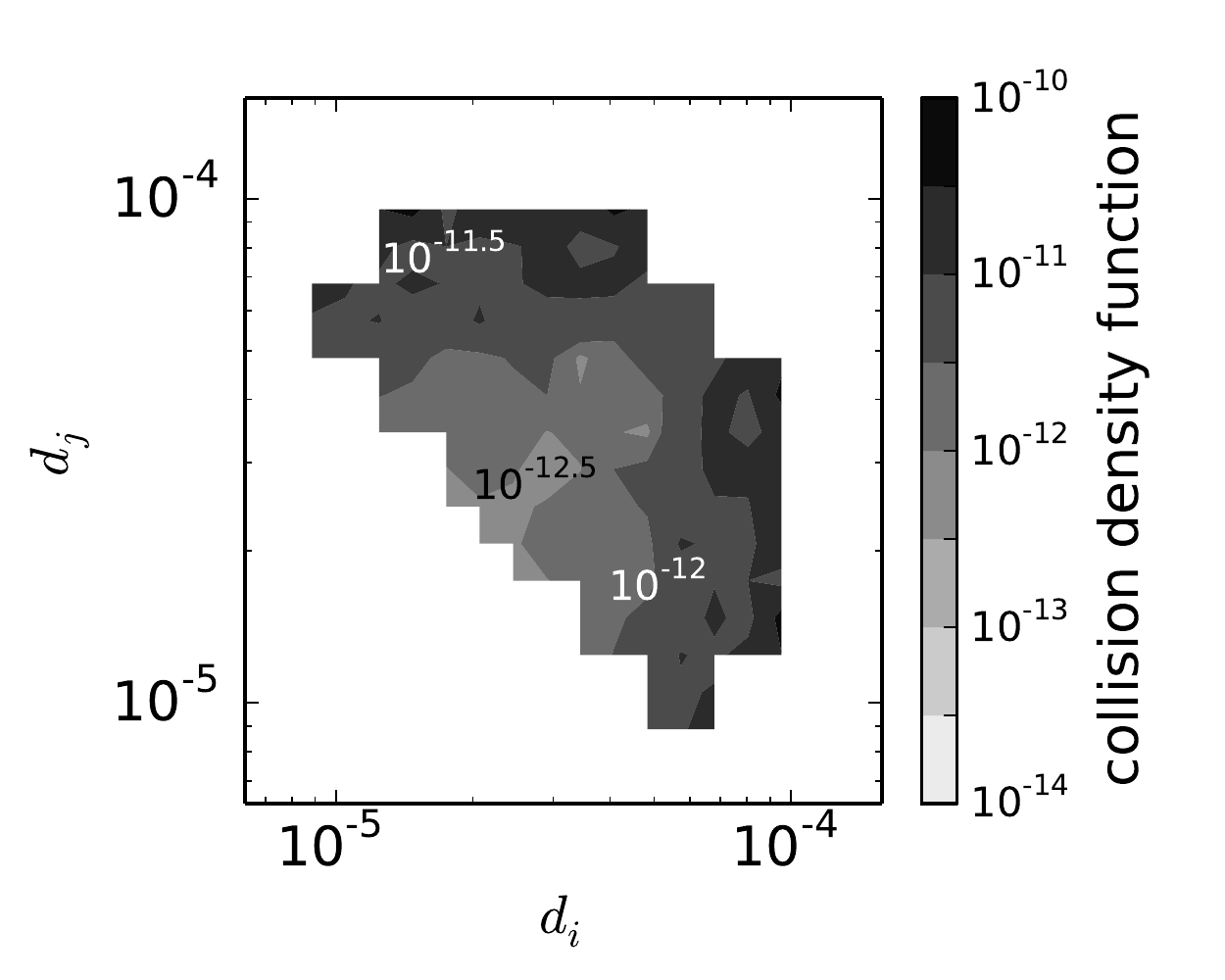}
  }
  \subfloat[initial solid volume fraction $\varepsilon_{s,0} = 0.16$ (Poly04)]{
  \includegraphics[width=0.45\textwidth]{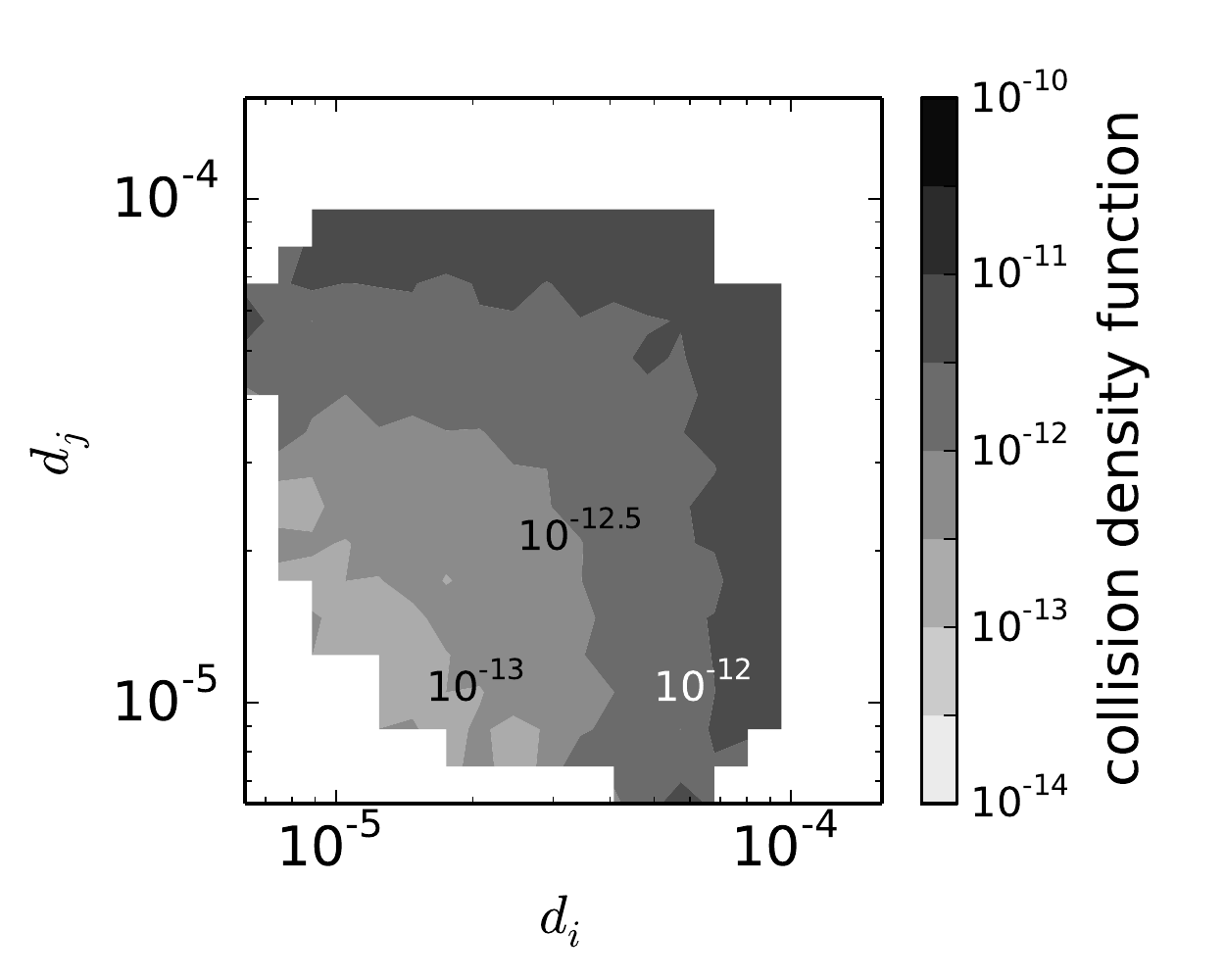}
  }
  \vspace{0.1in}
  \subfloat[binary collision theory based on Eq.~(\ref{eq:cff})]{
  \includegraphics[width=0.45\textwidth]{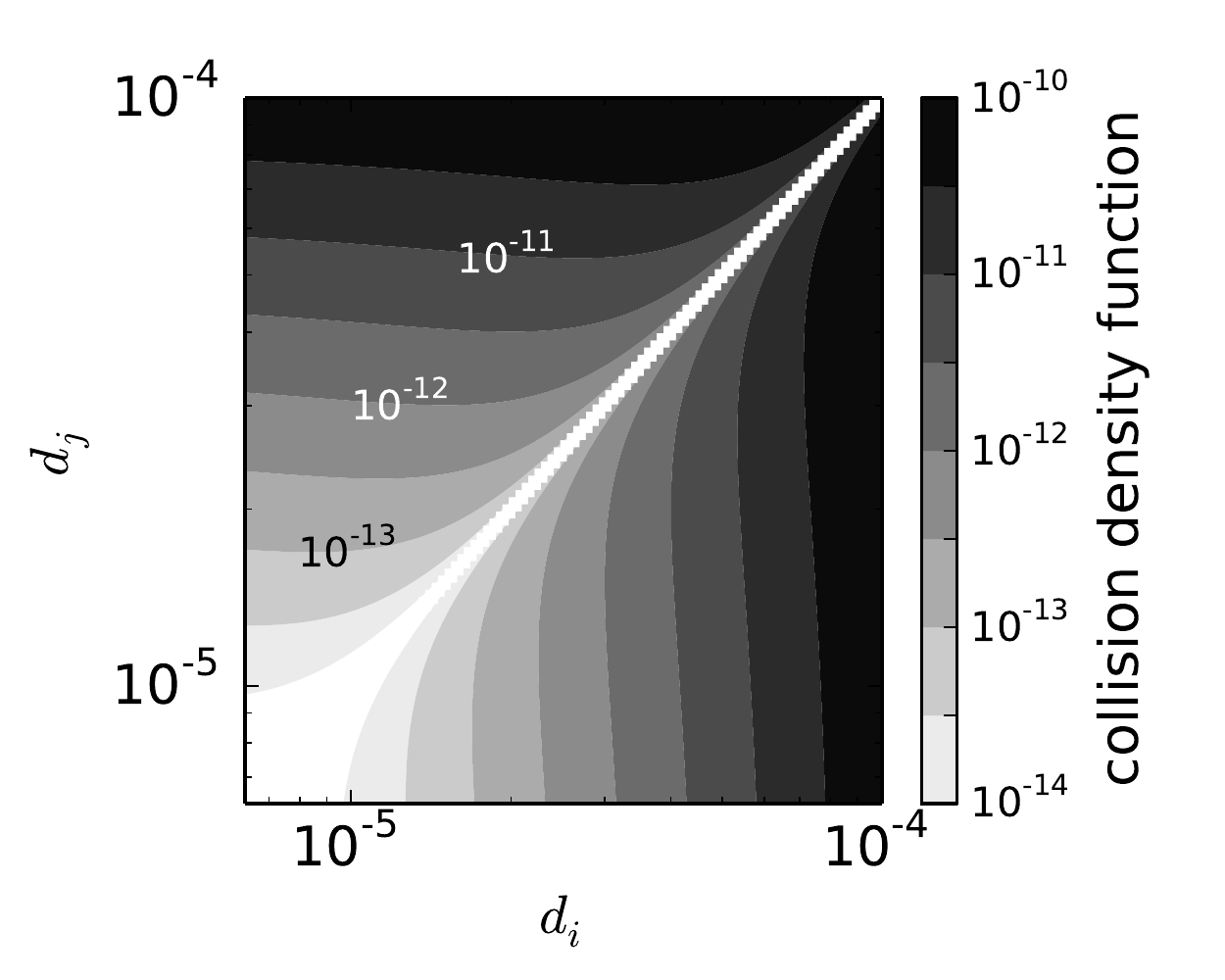}
  }
  \caption{Comparison of the collision frequency function obtained by binary collision theory and
  the results obtained in CFD--DEM simulations. The simulations are performed at Bond number Bo =
  2.}
  \label{fig:cff-comp}
\end{figure}

The results obtained at different $\varepsilon_{s,0}$ in the CFD--DEM simulations are compared with
those obtained by using Eq.~(\ref{eq:cff}) to demonstrate the variation of the collision frequency
functions due to solid volume fractions. The contours of collision frequency functions obtained in
the present simulations are shown in Figs.~\ref{fig:cff-comp}(a)~and~\ref{fig:cff-comp}(b), and the
contour computed based on binary collision theory is presented in Fig.~\ref{fig:cff-comp}(c). The
collision frequency functions are obtained at $t^* = 5.4$ in the CFD--DEM simulations before the
sedimentation of larger particles. It can be seen that the collision frequency function obtained at
low initial solid volume fraction is in good agreement with the theoretical solution in terms of the
order of magnitude.  However, the contour obtained at higher initial solid volume fraction
$\varepsilon_{s,0}$ is different from that obtained at lower $\varepsilon_{s,0}$.  Compared with
Fig.~\ref{fig:cff-comp}(a), it can be seen in Fig.~\ref{fig:cff-comp}(b) that the collision occurs
for smaller particles can be observed at higher initial solid volume fraction. When the initial
solid volume fraction increases, the average distance between smaller particles decreases, and thus
the flocculation between smaller particles is more likely to be observed. However, the collision
frequency between larger particles decreases at higher initial solid volume fraction. This is due to
the fact that smaller particles are filling the spaces of larger ones and reduces the chance of
direct contact between larger particles.

In summary, the collision frequency function obtained at low solid volume fraction
($\varepsilon_{s,0} = 0.04$) is in good agreement with the theoretical solution in terms of the
order of magnitude. However, a decrease of collision frequency is observed at high solid volume
fraction ($\varepsilon_{s,0} = 0.16$). This indicates the decrease of collision frequency at high
volume fractions in macro-scale modeling of cohesive silt sedimentation should be taken into
account.

\section{Conclusion}

In this work, we have studied the three-dimensional settling process of cohesive silt via numerical
simulations. The CFD--DEM approach is adopted in the simulations because it can faithfully represent
the individual and averaged quantities of sediment particles. The van der Waals force is taken into
account to describe the cohesive forces between silt particles. In addition, the different behaviors
of both mono- and poly-dispersed silt in the settling process are investigated.  Phenomena observed
in the experimental study~\citep{te13sp,te15hindered}, including the decrease of structural
densities, the segregation, and the variation of settling velocity, are all captured in the
numerical simulations.  It is demonstrated that the structural densities of the sediment beds
decrease significantly when the particle Bond number increases. This is because the increase of
particle cohesion between the silt particles prevents the formation of more close-packed structures.
Moreover, the mean structural density decreases further when the polydispersity of particle size
increases. This is attributed to the fact that cohesive force is larger for smaller particles at a
given Hamaker coefficient. The particle cohesion can also reduce the segregation of the
characteristic lines at high solid volume fraction ($\varepsilon_{s,0} = 0.16$) because smaller
particles are overtaken by larger ones. The force ratio obtained by using silt particle is also
presented to demonstrate its relationship with the packing arrangement. The force ratios obtained in
the present study is consistent with those results using much larger particles in the literature. 

In addition to the validation against existing data in the literature, the micromechanics of
individual silt particle during the settling process is investigated. The cohesive force network
between silt particles is presented to demonstrate the propagation of the cohesive force. At lower
initial solid volume fraction, small and isolated flocs are observed. On the other hand, the number
of contacts between silt particles increases significantly at higher initial solid volume fraction.
The collision frequency function obtained in the present simulations is compared with the formula in
the literature. A decrease in the collision frequency function is observed at high initial solid
volume fraction ($\varepsilon_{s,0} = 0.16$). This effect should be accounted for in the modeling of
cohesive silt sedimentation using empirical models for macro-scale problems.

\section*{Acknowledgments}

The computational resources used for this project were provided by the Advanced Research Computing
(ARC) of Virginia Tech, which is gratefully acknowledged. RS gratefully acknowledges Dr. Prashant
Gupta for the implementation and validation of the inter-particle cohesive force module of
\textit{SediFoam}.

\section*{Reference}
\bibliographystyle{elsarticle-harv}
\bibliography{references}

\appendix
\setcounter{secnumdepth}{0}
\section{Appendix: Generation of Silt Particles Based on Material Weight Fraction}

\renewcommand\thefigure{A.\arabic{figure}}
\renewcommand\theequation{A.\arabic{equation}}
\setcounter{figure}{0}

The fraction of silt by material weight is approximated by using log-normal distribution according
to the literature~\citep{vanoni06se,garcia08se}.  However, the material weight fraction cannot be
directly used to generate the poly-dispersed particles in CFD--DEM. This is because the number of
particles per logarithmic interval is unknown. Instead, particle number distribution is required to
generate particles of random diameters. According to~\cite{heintzenberg94pl}, when the distribution
of material weight is log-normal, the distribution of particle number is also log-normal. The number
median particle diameter $d_{n50} = d_{50}\exp(-3\sigma^2)$, where $\sigma$ is the variance of the
material weight fraction; the variance of particle diameter distribution $\sigma_{n}$ is consistent
with $\sigma$.

\begin{figure}[htbp]
  \centering
  \subfloat[percentage finer by material weight]{
  \includegraphics[width=0.45\textwidth]{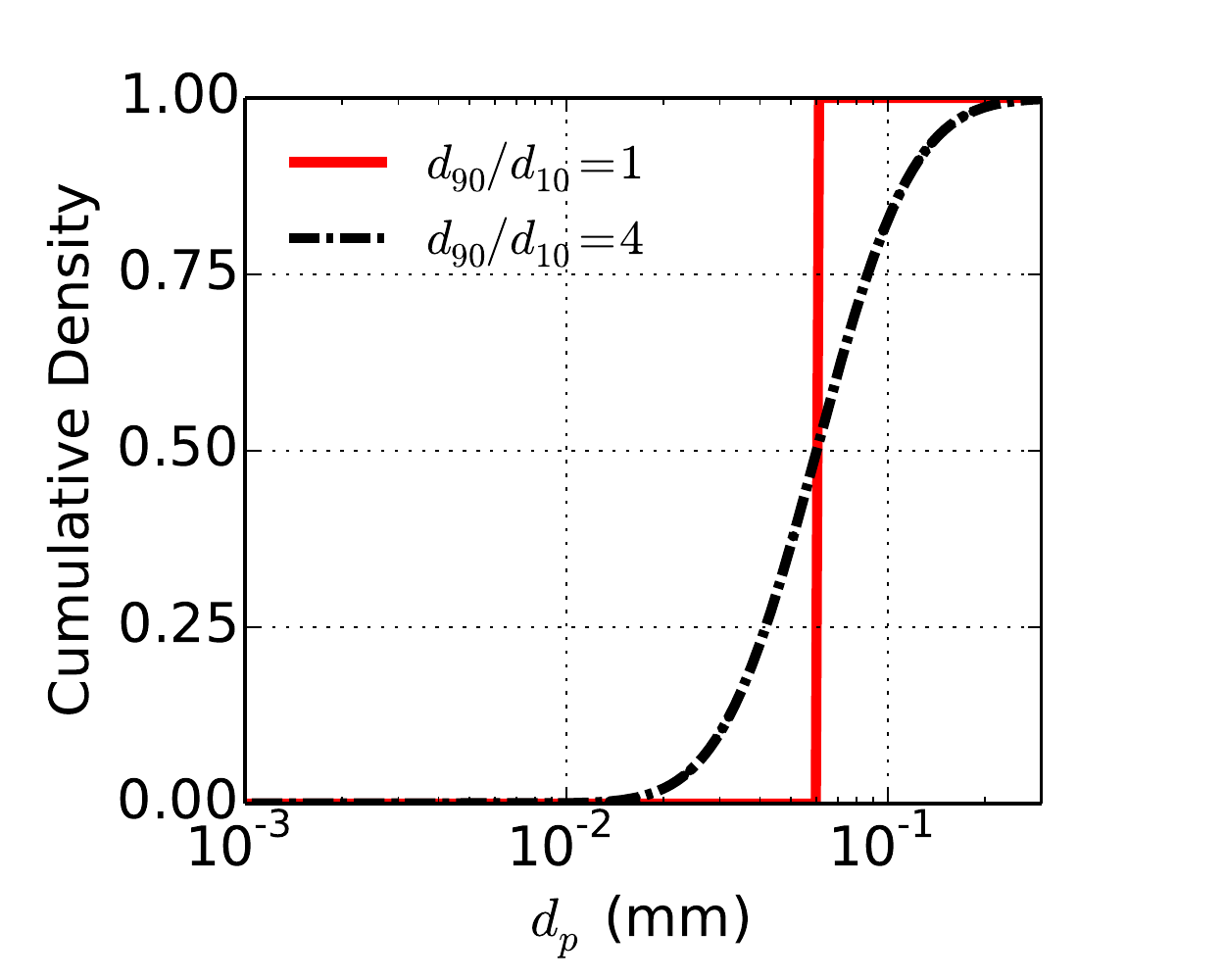}
  }
  \subfloat[probability density function of particle number]{
  \includegraphics[width=0.45\textwidth]{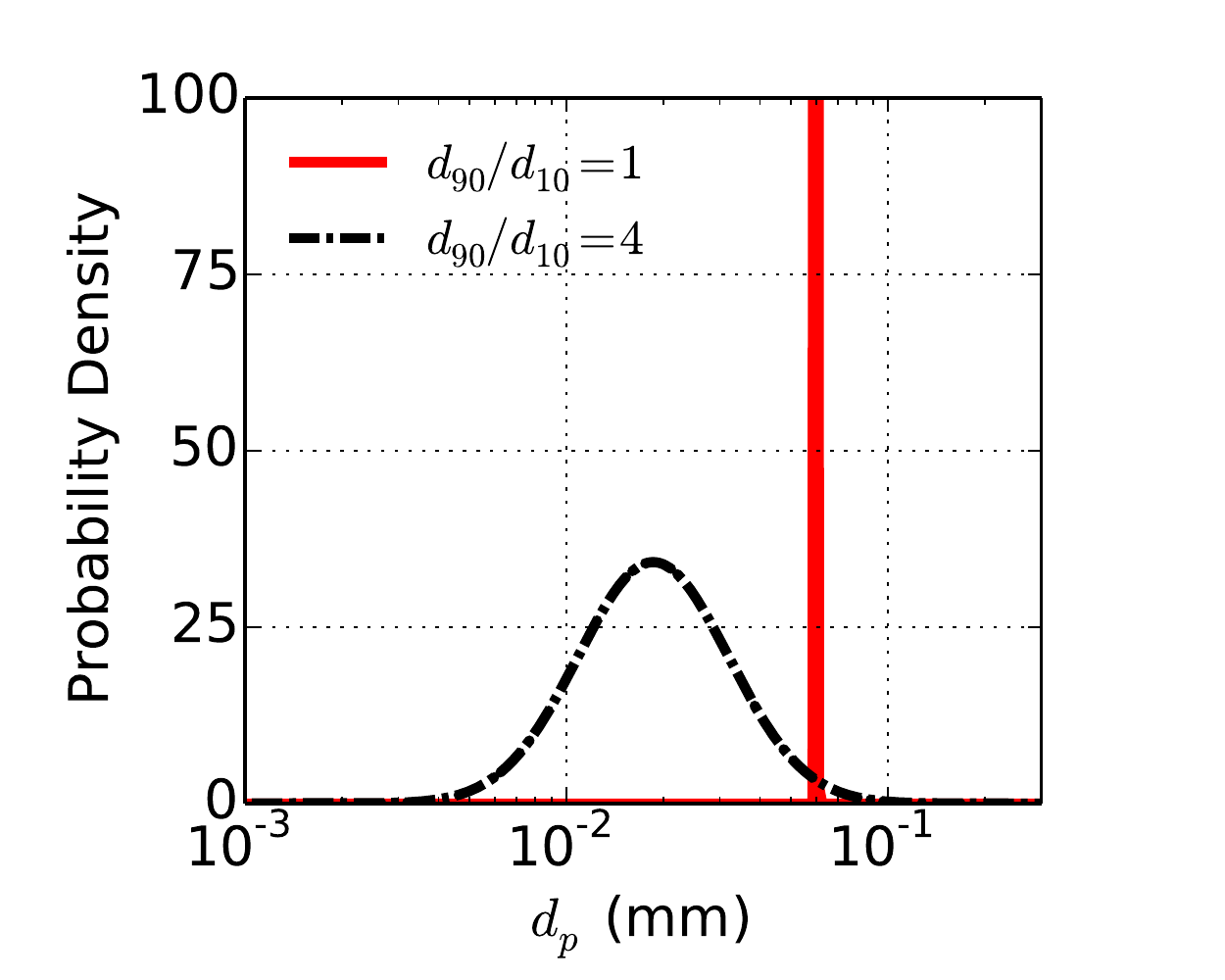}
  }
  \caption{Comparison of the distributions between mono- and poly-dispersed particles used in the
    simulations. In Panel (a), the cumulative density function of material weight is plotted as a
    function of particle size; in Panel (b), the distribution of particle number is plotted.}
  \label{fig:particleSize}
\end{figure}

The material weight distribution and the particle number distribution used in the simulations are
shown in Figs.~\ref{fig:particleSize}(a)~and~\ref{fig:particleSize}(b), respectively.  It can be seen
in Fig.~\ref{fig:particleSize}(a) that the mean particle diameters $d_{50}$ are identical for both
mono- and poly-dispersed silt.  Although the mean particle diameters are the same for both mono- and
poly-dispersed silt, most silt particles are smaller than mean diameter $d_{50}$ when the particle
is poly-dispersed ($d_{90}/d_{10} = 4$).  This is because a larger number of small particles are
required for the same material mass. 

%


%
%
\end{document}